\documentstyle[]{mn}
\newif\ifAMStwofonts

\ifoldfss
  \newcommand{\rmn}[1] {{\rm #1}}

  \ifCUPmtlplainloaded \else
    \NewTextAlphabet{textbfit} {cmbxti10} {}
    \NewTextAlphabet{textbfss} {cmssbx10} {}
    \NewMathAlphabet{mathbfit} {cmbxti10} {} 
    \NewMathAlphabet{mathbfss} {cmssbx10} {} 
  \fi
  \ifAMStwofonts
    \ifCUPmtlplainloaded \else
      \NewSymbolFont{upmath} {eurm10}
      \NewSymbolFont{AMSa} {msam10}
      \NewMathSymbol{\upi}     {0}{upmath}{19}
      \NewMathSymbol{\umu}     {0}{upmath}{16}
      \NewMathSymbol{\upartial}{0}{upmath}{40}
      \NewMathSymbol{\leqslant}{3}{AMSa}{36}
      \NewMathSymbol{\geqslant}{3}{AMSa}{3E}

      \let\leq=\leqslant 
      \let\geq=\geqslant 
    \fi
  \fi
\fi 

\ifnfssone
  \newmathalphabet{\mathit}
  \addtoversion{normal}{\mathit}{cmr}{m}{it}
  \addtoversion{bold}{\mathit}{cmr}{bx}{it}
  \newcommand{\rmn}[1] {\mathrm{#1}}

  \newmathalphabet{\mathbfit} 
  \addtoversion{normal}{\mathbfit}{cmr}{bx}{it}
  \addtoversion{bold}{\mathbfit}{cmr}{bx}{it}
  \newmathalphabet{\mathbfss} 
  \addtoversion{normal}{\mathbfss}{cmss}{bx}{n}
  \addtoversion{bold}{\mathbfss}{cmss}{bx}{n}
  \ifAMStwofonts
    \ifCUPmtlplainloaded \else
      \UseAMStwoboldmath
      \makeatletter
      \new@mathgroup\upmath@group
      \define@mathgroup\mv@normal\upmath@group{eur}{m}{n}
      \define@mathgroup\mv@bold\upmath@group{eur}{b}{n}
      \edef\UPM{\hexnumber\upmath@group}
      \new@mathgroup\amsa@group
      \define@mathgroup\mv@normal\amsa@group{msa}{m}{n}
      \define@mathgroup\mv@bold\amsa@group{msa}{m}{n}
      \edef\AMSa{\hexnumber\amsa@group}
      \makeatother
      \mathchardef\upi="0\UPM19
      \mathchardef\umu="0\UPM16
      \mathchardef\upartial="0\UPM40
      \mathchardef\leqslant="3\AMSa36
      \mathchardef\geqslant="3\AMSa3E

      \let\leq=\leqslant 
      \let\geq=\geqslant 
    \fi
  \fi
\fi 

\ifnfsstwo
  \newcommand{\rmn}[1] {\mathrm{#1}}

  \DeclareMathAlphabet{\mathbfit}{OT1}{cmr}{bx}{it}
  \SetMathAlphabet\mathbfit{bold}{OT1}{cmr}{bx}{it}
  \DeclareMathAlphabet{\mathbfss}{OT1}{cmss}{bx}{n}
  \SetMathAlphabet\mathbfss{bold}{OT1}{cmss}{bx}{n}
  \ifAMStwofonts
    \ifCUPmtlplainloaded \else
      \DeclareSymbolFont{UPM}{U}{eur}{m}{n}
      \SetSymbolFont{UPM}{bold}{U}{eur}{b}{n}
      \DeclareSymbolFont{AMSa}{U}{msa}{m}{n}
      \DeclareMathSymbol{\upi}{0}{UPM}{"19}
      \DeclareMathSymbol{\umu}{0}{UPM}{"16}
      \DeclareMathSymbol{\upartial}{0}{UPM}{"40}
      \DeclareMathSymbol{\leqslant}{3}{AMSa}{"36}
      \DeclareMathSymbol{\geqslant}{3}{AMSa}{"3E}

      \let\leq=\leqslant 
      \let\geq=\geqslant 
    \fi
  \fi
\fi 

\ifCUPmtlplainloaded \else
  \ifAMStwofonts \else 
    \def\upi{\pi}
    \def\umu{\mu}
    \def\upartial{\partial}
  \fi
\fi

\newcommand{\COBE}
        {{\em COBE}}

\newcommand{\MAP}
	{{\em MAP}}
\newcommand{\Planck}
	{{\em Planck}}
\newcommand{\Boomerang}
	{BOOMERanG}
\newcommand{\CAT}
	{CAT}

\newcommand{\CBI}
	{CBI}
\newcommand{\DASI}
	{DASI}

\newcommand{\lapack}
	{{\sc lapack}}

\newcommand{\cmb}
        {CMB}

\newcommand{\samethanks}
	{{\Huge $^\star$}}
\newcommand{\order}
	{O}

\newcommand{\smooth}[1]
	{\bar{#1}}
\newcommand{\vect}[1]
        {\mbox{\boldmath ${#1}$}}
\newcommand{\matr}[1]
	{\mbox{\bf \sf{#1}}}

\newcommand{\skyhat}
	{\hat{\vect{r}}}

\newcommand{\etc}
	{etc.}
\newcommand{\etal}
	{et al.}
\newcommand{\eg}
	{e.g.}
\newcommand{\cf}
	{cf.}
\newcommand{\ie}
	{i.e.}
\newcommand{\eq}[1]
	{equation~(\ref{equation:#1})}
\newcommand{\eqs}[1]
	{equations~(\ref{equation:#1})}
\newcommand{\Eq}[1]
        {Equation~(\ref{equation:#1})}

\newcommand{\sect}[1]
	{Section~\ref{section:#1}}
\newcommand{\apdx}[1]
        {{\mbox Appendix~\ref{section:#1}}}

\newcommand{\sects}[1]
        {Sections~\ref{section:#1}}

\newcommand{\tabl}[1]
        {{\mbox Table~\ref{table:#1}}}

\newcommand{\fig}[1]
	{Fig.~\ref{figure:#1}}

\newlength{\singlefigureheight}
\newlength{\doublefigureheight}
\newlength{\triplefigureheight}
\newlength{\squarefigureheight}
\newcommand{\AaA}
        {A\&A}
\newcommand{\AaAS}
        {A\&AS}

\newcommand{\ApJ}
        {ApJ}

\newcommand{\MNRAS}
        {MNRAS}

\newcommand{\Nature}
        {Nature}
\newcommand{\AstSpace}
	{Ast.\ \& Space Sci.}

\newcommand{\PhysRep}
        {Phys.\ Rep.}

\newcommand{\PRD}
        {Phys.\ Rev.\ D}

\newcommand{\NewAstRev}
	{New Ast.\ Rev.}

\newcommand{\lmax}
	{l_{\rmn max}}

\newcommand{\svd}
	{SVD}

\newcommand{\FL}
        {FL}

\newcommand{\gr}
	{GR}
\newcommand{\Gr}
	{GR}
\newcommand{\sphere}
	{S}
\newcommand{\cutsphere}
	{{S^\prime}}
\newcommand{\mathword}[1]
{\,\,\,\,\,\,\, {\rmn{#1}} \,\,\,\,\,\,\,}
\newcommand{\integer}
	{{\rmn int}}

\begin{document}

\title[Incomplete sky \cmb\ analysis]
{Analysis of cosmic microwave background data on an incomplete sky}

\author[D.\ J.\ Mortlock, A.\ D.\ Challinor and M.\ P.\ Hobson]
       {
        Daniel J.\ Mortlock,$^{1,2}$\thanks{
		E-mail: mortlock@ast.cam.ac.uk (DJM);
		a.d.challinor@mrao. cam.ac.uk (ADC);
		mph@mrao.cam.ac.uk (MPH)}
	Anthony D.\ Challinor$^1$\samethanks\ 
	and Michael P.\ Hobson$^1$\samethanks\ \\
        $^1$Astrophysics Group, Cavendish Laboratory, Madingley Road,
        Cambridge CB3 0HE, U.K. \\
	$^2$Institute of Astronomy, Madingley Road, Cambridge
	CB3 0HA, U.K. \\
       }

\date{
Accepted 2001 October 24. 
Received 2001 June 7; in original form 2000 August 7}

\pagerange{\pageref{firstpage}--\pageref{lastpage}}
\pubyear{2001}

\label{firstpage}

\maketitle

\begin{abstract}
Measurement of the angular power spectrum of the cosmic microwave 
background is most often based on a spherical harmonic analysis of 
the observed temperature anisotropies. 
Even if all-sky maps are obtained, however,
it is likely that the region around 
the Galactic plane will have to be removed due to its strong microwave 
emissions.
The spherical harmonics are not orthogonal on the cut sky, but an 
orthonormal basis set can be constructed from 
a linear combination of the original functions. 
Previous implementations of this technique,
based on Gram-Schmidt orthogonalisation,
were limited to maximum 
Legendre multipoles of $\lmax \la 50$ as they required all the modes
have appreciable support on the cut sky, whereas for large $\lmax$
the fraction of modes supported is equal to the fractional area of the
region retained.
This problem is solved by using a singular value decomposition to remove 
the poorly-supported basis functions, although the treatment of the 
non-cosmological monopole and dipole modes necessarily becomes more 
complicated.
A further difficulty is posed by computational limitations -- 
orthogonalisation for a general cut requires $O(\lmax^6)$ operations and 
$O(\lmax^4)$ storage and so is impractical for $\lmax \ga 200$ at present.
These problems are circumvented for the special case of constant (Galactic) 
latitude cuts, for which the storage requirements 
scale as $O(\lmax^2)$ and the 
operations count scales as $O(\lmax^4)$.
Less clear, however, is the stage of the data analysis at which
the cut is best applied. As convolution is ill-defined on the incomplete
sphere, beam-deconvolution should not be performed after the cut, and,
if all-sky component separation is as successful as simulations indicate,
the Galactic plane should probably be removed immediately prior to
power spectrum estimation.
\end{abstract}


\section{Introduction}
\label{section:intro}

Since the first measurements of the temperature anisotropy of 
the cosmic microwave background (\cmb) by the 
{\em Cosmic Background Explorer} (\COBE) satellite (Smoot \etal\ 1992), 
a number of sophisticated experiments
have been undertaken to
measure the fluctuations at higher resolutions and sensitivities
(\eg\
Scott \etal\ 1996; Tanaka \etal\ 1996; 
Netterfield \etal\ 1997;
de Oliveira-Costa \etal\ 1998; 
Coble \etal\ 1999; 
de Bernardis \etal\ 2000; Wilson \etal\ 2000; Padin \etal\ 2001;
Halverson \etal\ 2001; Lee \etal\ 2001; Netterfield \etal\ 2001).
The primary result of these experiments has been 
the measurement of the angular power spectrum of the \cmb\ 
to Legendre multipoles of up to $l \simeq 1000$, which places strong
constraints on a number of cosmological parameters
(Lineweaver 1998; Efstathiou \etal\ 1999; de Bernardis \etal\ 2000;
Netterfield \etal\ 2001;
Wang, Tegmark \& Zaldarriaga 2001
and references therein).
In the future the 
{\em Microwave Anisotropy Probe} (\MAP; \eg\ Jarosik \etal\ 1998)
and the \Planck\ satellite (\eg\ Bersanelli \etal\ 1996)
will produce 
maps of the microwave sky with resolutions of between 5 and 30 arcmin
at a number of frequencies. 
Such extraordinary data-sets, consisting of millions of 
independent measurements, will clearly require novel analysis techniques.

One of the many difficulties is the treatment of the non-cosmological
contributions to the observed microwave sky.
Dust, synchrotron and free-free 
emission from the Galaxy (\eg\ Haslam \etal\ 1982; 
Schlegel, Finkbinder \& Davies 1998);
radio galaxies and other extra-Galactic `point' sources 
(\eg\ Toffolatti \etal\ 1998); and  
the Sunyaev-Zel'dovich (1970) effect caused by 
galaxy clusters (\eg\ Birkinshaw 1999)
all obscure the \cmb\ at some level
(see Hu, Sugi\-yama \& Silk 1997 or Barreiro 2000 for more 
complete reviews),
although these components have quite distinct spectral properties
and so can be separated using 
multi-frequency observations
(\eg\ Bennett \etal\ 1992;
Tegmark \& Efstathiou 1996;
Hobson \etal\ 1998; 
Bouchet \& Gispert 1999;
Jones, Hobson \& Lasenby 1999; Baccigalupi \etal\ 2000).
However these techniques are not likely to be able to
extract the Galactic emissions completely
(Stolyarov \etal\ 2001),
leaving the removal of the Galactic plane as the only option.
The Galaxy contributes relatively little at high latitudes
(\eg\ Haslam \etal\ 1982; Schlegel \etal\ 1998) so 
this is an acceptable, if not optimal, solution.
For instance, G\'{o}rski \etal\ (1994) removed the band 
within 20 deg of the Galactic plane to estimate the power spectrum
of the two-year \COBE\ Differential Microwave Radiometer sky
maps, and similar cuts have been proposed by both the \MAP\ and
\Planck\ collaborations. 
An essentially equivalent problem is posed if the survey's 
sky coverage is incomplete,
although there is less choice about 
the geometry of the cut in this case.

A number of aspects of the analysis become more difficult on
an incomplete sphere, one
of the most obvious reasons being that the spherical harmonics 
are no longer an orthonormal basis set. 
The most successful component separation techniques to date 
(\eg\ Hobson \etal\ 1998; Bouchet \& Gispert 1999)
rely on a mode-by-mode analysis which explicitly utilises the 
orthogonality of the spherical harmonics, 
although it may be preferable to 
remove the Galactic plane only
when estimating the \cmb\ power spectrum.
Unbiased power spectrum estimation using the spherical harmonics
is possible on the cut sky
(Wandelt, G\'{o}rski \& Hivon 2001),
but the covariance structure of the resulting psuedo-harmonics
is far from ideal,
so analysis using an orthonormal basis set is preferable. 
In particular the noise covariance matrix remains diagonal
in the case of spatially uniform (and uncorrelated) noise. 

It is possible to construct an orthonormal basis set from 
linear combinations of the spherical harmonics,
and an elegant implementation of this,
based on Cholesky decomposition of the 
coupling matrix of the spherical harmonics
on the cut sky,
was described 
by G\'{o}rski (1994). 
However the coupling matrix becomes ill-conditioned
for $\lmax \ga 50$, and so this method cannot 
be used to perform cut-sky orthogonalisation 
for either the \MAP\ experiment (with $\lmax \simeq 1500$) 
or the \Planck\ mission (with $\lmax \simeq 2500$).

A general formalism for orthogonalisation of the spherical
harmonics is presented in \sect{orthog},
although implementation to high $\lmax$ is only
possible at present in the special case of a constant 
latitude cut. 
The relationship between the various harmonic coefficients 
is discussed in \sect{harmonics},
and the extension of these results to \cmb\ analysis techniques
(specifically component separation and power spectrum estimation)
is covered in \sect{anal}.
The results are summarised and future possibilities 
are discussed in \sect{concs}.
Finally, 
the chosen conventions for the spherical 
harmonics are defined in \apdx{sphar}; 
formul\ae\ for integrals of the products of Legendre 
functions are given in \apdx{integ};
and the treatment of the non-cosmological monopole and dipole 
modes are discussed in \apdx{noncosmol}.

\section{Orthogonalisation of scalar basis\\ 
	functions}
\label{section:orthog}

The physics of the \cmb\ is most naturally expressed in Fourier
space, and it is standard practice to represent sky maps by 
their harmonic coefficients.
The basis functions chosen here are the real spherical harmonics,
$Y_{l,m}(\skyhat)$ (as defined in \apdx{sphar}), which form an
orthonormal basis on the complete sphere, $\sphere$.
In general $l \geq 0$ and $- l \leq m \leq l$, although
in practice a finite $\lmax$ must be used, which implies a
band-limit.
It is convenient to combine the two
indices, allowing the basis set to be expressed as a vector,
$\vect{Y}(\skyhat) = [Y_1(\skyhat), Y_2(\skyhat), \ldots,
Y_{i_{\rmn max}}(\skyhat)]^{\rmn T}$, where $i_{\rmn max} = (\lmax + 1)^2$.
There are several reasonable choices for the indexing,
$i(l, m)$, most notably grouping coefficients in $l$ or $m$,
as defined in \apdx{sphar}. Grouping in $l$ is most
natural for power spectrum estimation, but grouping in $m$
is more efficient computationally in cases of azimuthal symmetry
(\sect{constlat}).

The spherical harmonics are not orthogonal on the incomplete
sphere $\cutsphere$, 
as can be seen from the structure of their coupling matrix
(\sect{coupling}).
A decomposition of the coupling matrix can be used to construct
an orthonormal basis set (\sect{method}), but implementation to 
high-resolution is currently possible only in the special case of 
constant latitude cuts (\sect{constlat}).

\subsection{The coupling matrix}
\label{section:coupling}

The coupling matrix of a set of functions encodes 
their orthogonality and normalisation
properties over a given range.
In the case of the spherical harmonics on the incomplete
sphere it is given by
\begin{equation}
\label{equation:coupling}
\matr{C} = \int_{\cutsphere} \vect{Y}(\skyhat)
\vect{Y}^{\rmn T} (\skyhat) \, {\rmn d \Omega}.
\end{equation}
If $\cutsphere = \sphere$ then
the harmonics are orthonormal and
$\matr{C} = \matr{I}$; otherwise the off-diagonal elements
are non-zero, indicating that the basis functions are 
non-orthogonal.

An alternative formulation is to introduce a window function, 
$w(\skyhat)$, so that
\begin{equation}
\label{equation:coupling_smooth}
\matr{C} = \int_{\sphere} w^2(\skyhat) 
\vect{Y}(\skyhat)
\vect{Y}^{\rmn T} (\skyhat) \, {\rmn d \Omega}.
\end{equation}
In some ways this approach is more flexible, as $w(\skyhat)$ can either be a
smoothly-varying apodizing function (\cf\ Tegmark 1997)
or take the form
\begin{equation}
\label{equation:window}
w_{\cutsphere} (\skyhat) = 
\left\{
\begin{array}{lll}
1, & {\rmn if} & \skyhat \in \cutsphere, \\
& & \\
0, & {\rmn if} & \skyhat \in \sphere - \cutsphere,
\end{array}
\right.
\end{equation}
mimicing the effect of the sharp cut defined above. 
However this definition of the window function can lead
to inconsistencies if a band-limited analysis is carried out, as 
$w_{\cutsphere} (\skyhat)$ cannot be properly represented by a 
finite analysis (see \sect{highres_harm}).
It is for this reason that the first formalism is used here,
although most of the subsequent results can also be derived 
using window functions.

For a pixel-based analysis, the coupling matrix
can be defined by replacing the integral in \eq{coupling}
by a sum over points on the sphere (\ie\ pixel centres),
$\skyhat_p$, where $1 \leq p \leq N_{\rmn p}$ 
and $N_{\rmn p}$ is the number of pixels. In this
case 
\begin{equation}
\label{equation:coupling_pix}
\matr{C} = 
\sum_{p = 1; \skyhat_p \in \cutsphere}^{N_{\rmn p}} \vect{Y}(\skyhat_p)
\vect{Y}^{\rmn T} (\skyhat_p) \Omega_p,
\end{equation}
where $\Omega_p$ is the area of the $p$th pixel
and there is no pixel-smoothing (\cf\ G\'{o}rski 1994).
In the limit $N_{\rmn p} \rightarrow \infty$
\eqs{coupling} and (\ref{equation:coupling_pix}) become
equivalent and pixelisation issues become irrelevant.
If the points are
uniformly distributed over the sphere $\matr{C}$ should be close to
the identity, the small discrepancies merely reflecting the
approximation of the integral as a sum; otherwise $\matr{C}$
reflects the spatial distribution of the points 
much as in the continuum case,
but there is freedom to represent apodizing filters 
as well as discrete cuts.

Assuming $\Omega_\cutsphere > 0$, the coupling matrix
is formally symmetric, positive definite and invertible,
irrespective of which of the above definitions is used.
However $\matr{C}$ rapidly becomes numerically singular:
\eg\ if $\lmax = 50$,
the condition number\footnote{The condition number of a matrix is
the (absolute value of) the ratio of its greatest and smallest
eigenvalues; it is large for ill-conditioned matrices,
and infinite for singular matrices.} of $\matr{C}$
is $\sim 5 \times 10^9$
for a constant latitude cut of $b_{\rmn cut} = \pm 20$ deg.
This can be further understood in terms of the eigenstructure of
the coupling matrix.

\subsubsection{Eigenstructure}
\label{section:eigen}

\begin{figure*}
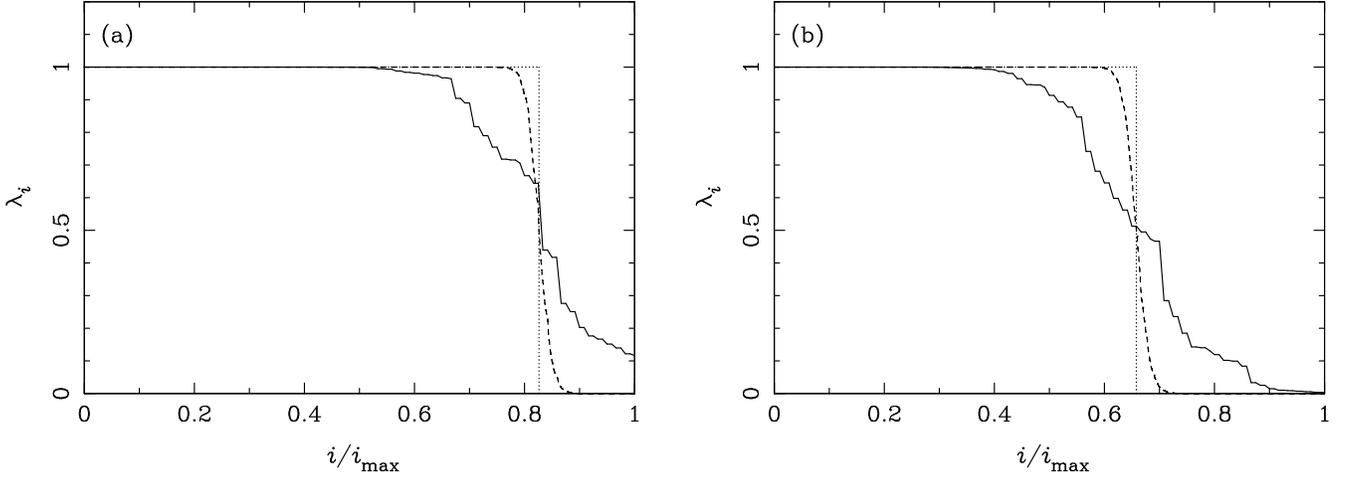

\includegraphics{evals10.ps}
\includegraphics{evals20.ps}
\vspace{\singlefigureheight}
\caption{The distribution of eigenvalues of $\matr{C}$, shown for
symmetric constant latitude cuts of $b_{\rmn cut} = 10$ deg in (a)
and $b_{\rmn cut} = 20$ deg in (b).
In both panels the eigenvalues are sorted into decreasing order,
and distributions are shown for
$\lmax = 10$ (solid line),
$\lmax = 100$ (dashed line)
and
$\lmax = 1000$ (dotted line).
The limiting cases of $\lmax \rightarrow \infty$
[in which $\Omega_{\cutsphere} / (4 \pi)$ of the eigenvalues
are unity and the remaineder zero]
are almost indistinguishable from the $\lmax = 1000$ distributions.}
\label{figure:eigenvalues}
\end{figure*}

The coupling matrix has $i_{\rmn max} = (\lmax + 1)^2$
eigenvectors, $\vect{v}_i$, and eigenvalues, $\lambda_i$,
which satisfy
\begin{equation}
\matr{C} \vect{v}_i = \lambda_i \vect{v}_i.
\end{equation}
Premultiplying by $\vect{Y}^{\rmn T} (\skyhat)$ and
expanding out the implicit summations gives
\begin{equation}
\label{equation:eigen_1}
\int_{\cutsphere} \sum_{k = 1}^{i_{\rmn max}}
Y_k(\skyhat^\prime) (v_i)_k
\sum_{j = 1}^{i_{\rmn max}}
Y_j(\skyhat) Y_j(\skyhat^\prime)
\, {\rmn d}\Omega^\prime
= \lambda_i v_i(\skyhat),
\end{equation}
where $v_i(\skyhat) = \vect{Y}^{\rmn T}(\skyhat) \vect{v}_i$ 
is the $i$th eigenfunction of the coupling matrix.
The completeness of the spherical harmonics 
in the $\lmax \rightarrow \infty$ limit
implies that $\vect{Y}^{\rmn T}(\skyhat) \vect{Y}(\skyhat^\prime)
= \delta(\skyhat - \skyhat^\prime)$ [where 
$\delta(x)$ is the Dirac delta function],
so that \eq{eigen_1}
reduces to
\begin{equation}
\label{equation:eval}
\int_{\cutsphere} v_i(\skyhat^\prime) \delta(\skyhat - \skyhat^\prime)
\, {\rmn d} \Omega^\prime
=
\lambda_i v_i(\skyhat).
\end{equation}
For a given $i$ this must be true at all $\skyhat$,
which implies that either:
$v_i(\skyhat) = 0$ in the cut region, $\sphere - \cutsphere$,
in which case $\lambda_i = 1$;
or
$v_i(\skyhat) = 0$ in $\cutsphere$,
in which case $\lambda_i = 0$.
In other words these eigenfunctions are completely localised in either the
cut sphere or the removed region.
This bimodality is only strictly true in the $\lmax \rightarrow \infty$
limit, but, as shown in \fig{eigenvalues}, is a good approximation
for $\lmax \ga 500$.

As the coupling matrix is symmetric, those eigenvectors with different
eigenvalues are orthogonal, and those with the same eigenvalues
can be made orthogonal by a rotation in the subspace defined by
the eigenvalue in question (\eg\ Arfken 1985). Thus the
eigenfunctions with $\lambda_i = 1$
represent an orthogonal basis set on $\cutsphere$,
whereas those with $\lambda_i = 0$ have no support in
this region and so cannot be orthogonal (or normalised) on the cut sky.
The freedom in choice of basis
does not extend to mixing the $\lambda_i = 0$ modes
(\ie\ those corresponding to the null-space of $\matr{C}$)
with the $\lambda_i = 1$ modes 
(\ie\ those in the range of $\matr{C}$), 
and so the number of supported modes is determined by a
combination of the band-limit and the cut.

The number of orthonormal basis functions (\ie\ the rank of \matr{C})
is proportional to the area of the sphere retained, 
$\Omega_\cutsphere$.
Hence it is possible to define only
\begin{equation}
\label{equation:rank}
i^\prime_{\rmn max} \simeq \frac{\Omega_\cutsphere}{\Omega_\sphere}
i_{\rmn max}
= 
\frac{\Omega_\cutsphere}{4 \pi}
i_{\rmn max}
\end{equation}
orthonormal functions on the cut sphere 
for a given (large) band-limit.
The relative reduction in the basis set is the same as would occur
in the equivalent pixel analysis: the number of pixels retained is also
given by
$N_{\rmn p}^\prime \simeq \Omega_\cutsphere / \Omega_\sphere \, N_{\rmn p}$.
For low $\lmax$ these arguments do not hold, and it is possible to 
create a basis set with more than 
$\Omega_\cutsphere / \Omega_\sphere \, i_{\rmn max}$ elements. 
Moreover, all these functions are required to ensure 
that the cut-sphere basis set is complete (as well as orthonormal)
in the case of a low band-limit.

\subsection{Construction of an orthonormal basis}
\label{section:method}

The construction of an orthonormal basis set from a set of 
linearly-independent functions is a well-established mathematical
technique, and a number of orthogonalisation methods are possible.
The most basic is Gram-Schmit orthogonalisation (\eg\ Arfken 1985), 
in which the new basis functions are built-up sequentially, but
this algorithm is numerically unstable. 
The modified Gram-Schmidt algorithm (\eg\ Golub \& van Loan 1996)
is stable, but it is generally preferable to use 
matrix techniques to create all the new 
basis functions simultaneously.

Starting with the spherical harmonics, $\vect{Y}(\skyhat)$,
the task is to find a set of functions
$\vect{Y}^\prime(\skyhat)$
which are orthonormal on the incomplete sphere.
In terms of a conversion matrix, $\matr{B}$, the two 
sets of functions 
are related by
\begin{equation}
\label{equation:yprime_general}
\vect{Y}^\prime(\skyhat) = \matr{B}\vect{Y}(\skyhat).
\end{equation}
Note that $\matr{B}$ has dimensions 
$i^\prime_{\rmn max} \times i_{\rmn max}$, 
where $i_{\rmn max} = (\lmax + 1)^2$ and 
$i^\prime_{\rmn max} \leq i_{\rmn max}$ is determined by the
band-limit and the cut, as described in \sect{coupling}.
It is also important to note that the indexing of the 
$Y^\prime_i(\skyhat)$ is qualitatively different from 
the $Y_i(\skyhat)$. The latter are really two-index objects, 
with their characteristic scale given by $\sim \pi / l$ 
and $m$ relating to `orientation'. However the new basis
functions include contributions from spherical harmonics
with different $l$-values, and thus do not have a well-defined
angular scale.
Hence their single index contains no physical information,
and the ordering or grouping of the new basis functions is arbitrary.

From \eq{coupling}, the coupling matrix of these new 
functions is 
\begin{eqnarray}
\label{equation:coupling_prime}
\matr{C}^\prime & = & \int_{\cutsphere} 
\vect{Y}^\prime (\skyhat)
{\vect{Y}^\prime}^{\rmn T} (\skyhat)
\, {\rmn d} \Omega \nonumber \\
& = & \matr{B} \matr{C} \matr{B}^{\rmn T}.
\end{eqnarray}
Hence any conversion matrix which satisfies\footnote{Here 
$\tilde{\matr{I}}$ is the $i^\prime_{\rmn max}
\times i^\prime_{\rmn max}$ identity matrix, as distinct 
from the (potentially larger) 
$i_{\rmn max} \times i_{\rmn max}$ identity matrix, $\matr{I}$.}
\begin{equation}
\label{equation:b_general}
\matr{B} \matr{C} \matr{B}^{\rmn T} = \tilde{\matr{I}}
\end{equation}
yields basis functions which are orthonormal on the 
cut sphere, and the task of orthogonalisation is reduced 
to finding a solution for $\matr{B}$ given $\matr{C}$.
Whilst such a solution does not exist for arbitrary $\matr{C}$,
in all cases of practical interest a suitable conversion matrix
can be constructed from the coupling matrix.
One possible method is 
direct calculation of the eigenstructure of $\matr{C}$,
which yields
a conversion matrix with elements given by
$B_{i^\prime, i} = (v_{i^\prime})_i \lambda_{i^\prime}^{-1/2}$, 
where the $\vect{v}_{i^\prime}$ are the eigenvectors of $\matr{C}$
and the $\lambda_{i^\prime}$ its positive eigenvalues.
However it is advantageous to include the symmetry of the
coupling matrix explicitly, which 
leads to a factorization of the form
\begin{equation}
\label{equation:coupling_decomp}
\matr{C} = \matr{A} \matr{A}^{\rmn T},
\end{equation}
where $\matr{A}$ is an 
$i_{\rmn max} \times i^\prime_{\rmn max}$ matrix,
the form of which 
is determined by the decomposition method.
Combining \eqs{b_general}
and (\ref{equation:coupling_decomp}),
the task of 
orthogonalisation is reduced to finding 
$\matr{B}$ such that 
\begin{equation}
\label{equation:b_orthog}
\matr{B} \matr{A} = \tilde{\matr{O}},
\end{equation}
where $\tilde{\matr{O}}$ is an $i^\prime_{\rmn max} \times
i^\prime_{\rmn max}$ orthogonal matrix (\ie\ 
$\tilde{\matr{O}} \tilde{\matr{O}}^{\rmn T}
= \tilde{\matr{O}}^{\rmn T} \tilde{\matr{O}} = \tilde{\matr{I}}$).

Whilst \eqs{coupling_decomp} and (\ref{equation:b_orthog}) 
are general expressions which must be satisfied by the
conversion matrix, they do not define a 
definite algorithm for the
orthogonalisation. 
In practice it is simplest to choose $\tilde{\matr{O}} = \tilde{\matr{I}}$,
leading to the requirement that 
\begin{equation}
\label{equation:b_decomp}
\matr{B} \matr{A} = \tilde{\matr{I}}. 
\end{equation}
However the optimal choice of decomposition method used to 
generate $\matr{A}$ depends on whether 
the coupling matrix is (numerically) invertible, and 
hence on the band-limit of the analysis.

\subsubsection{Low-resolution analysis}
\label{section:lowres_orthog}

\begin{figure*}
\includegraphics{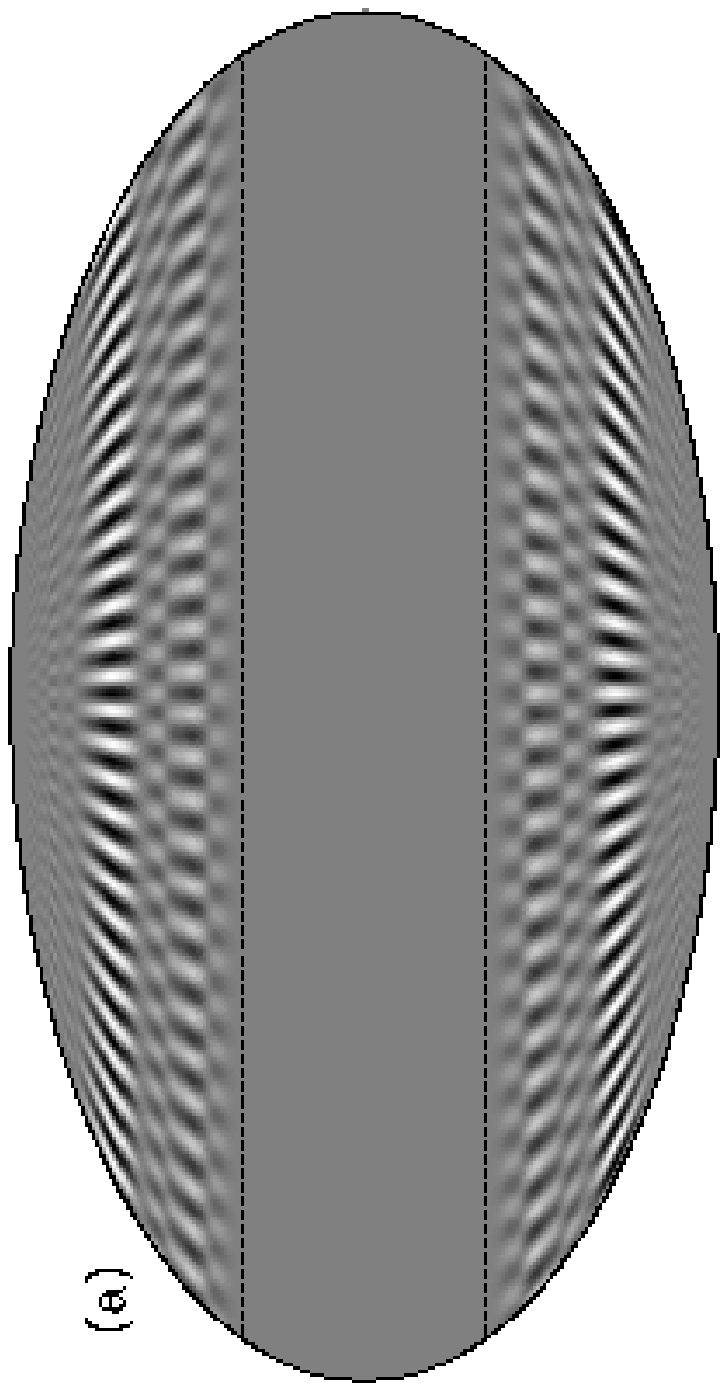}
\includegraphics{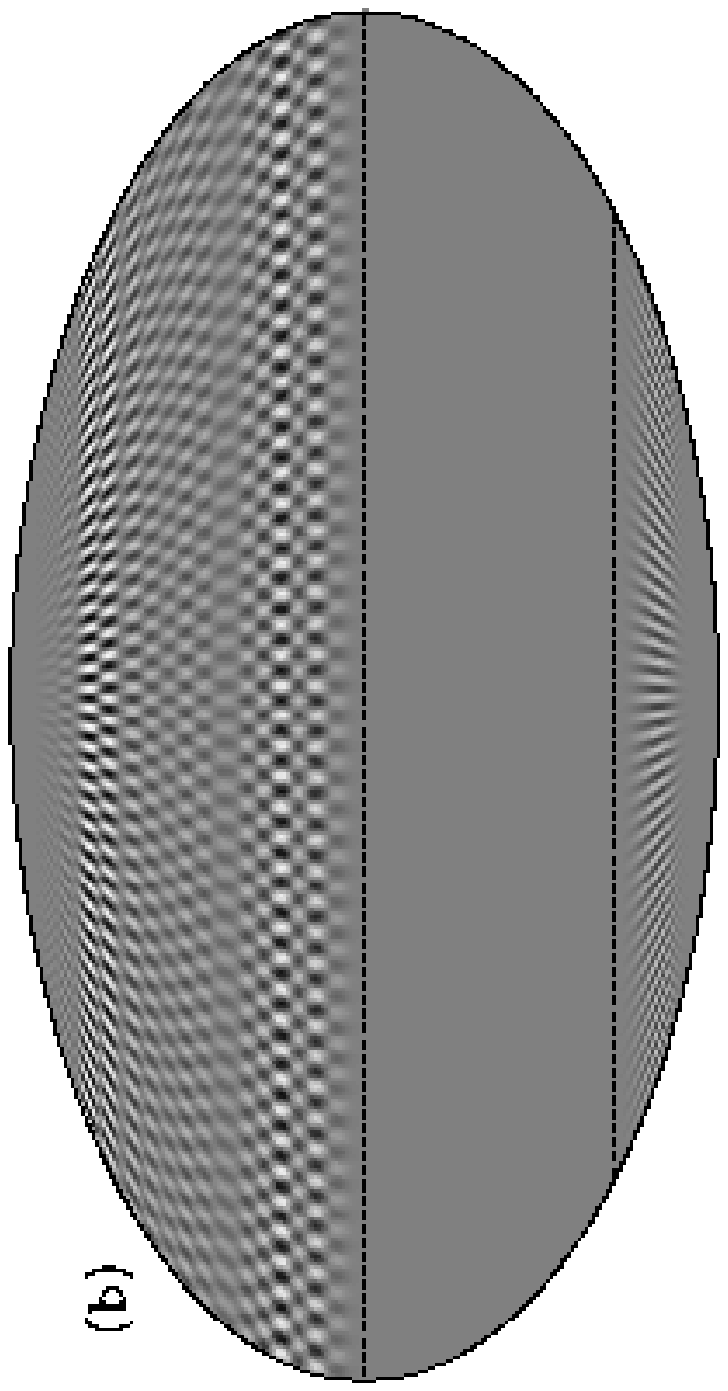}
\includegraphics{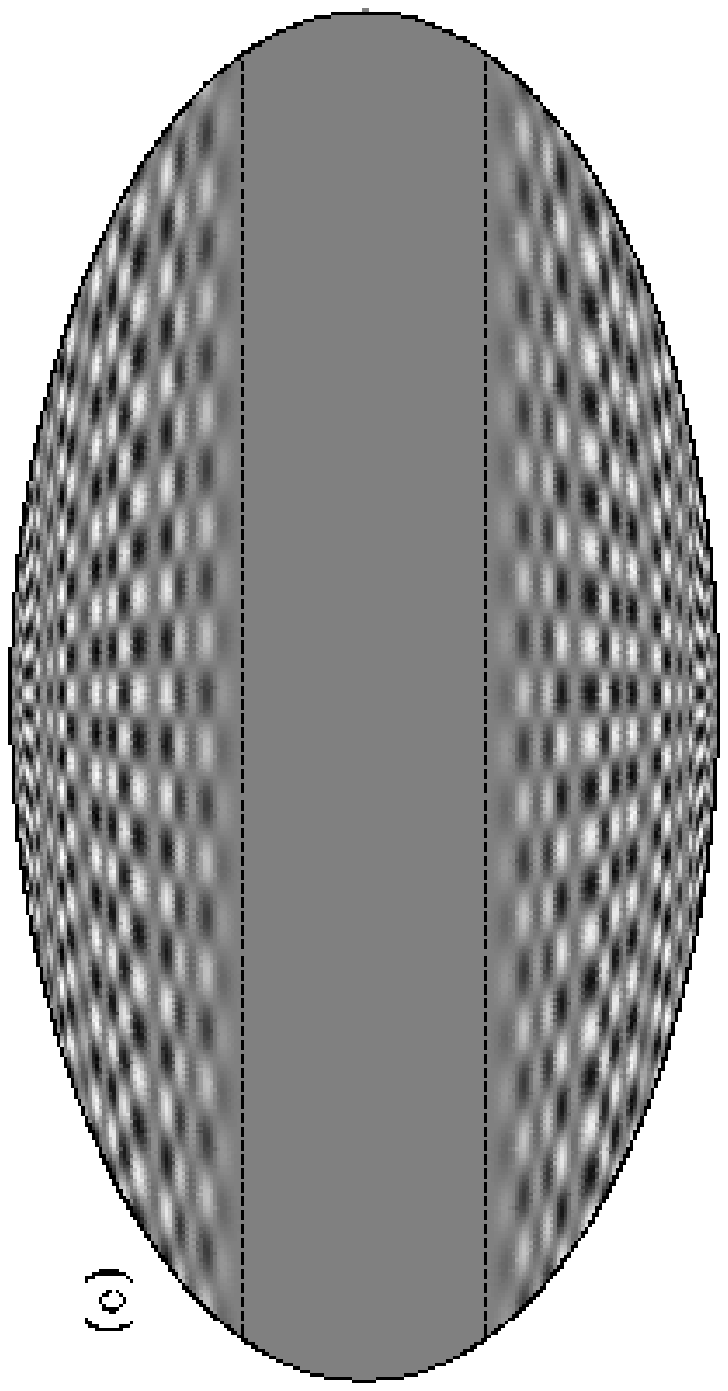}
\includegraphics{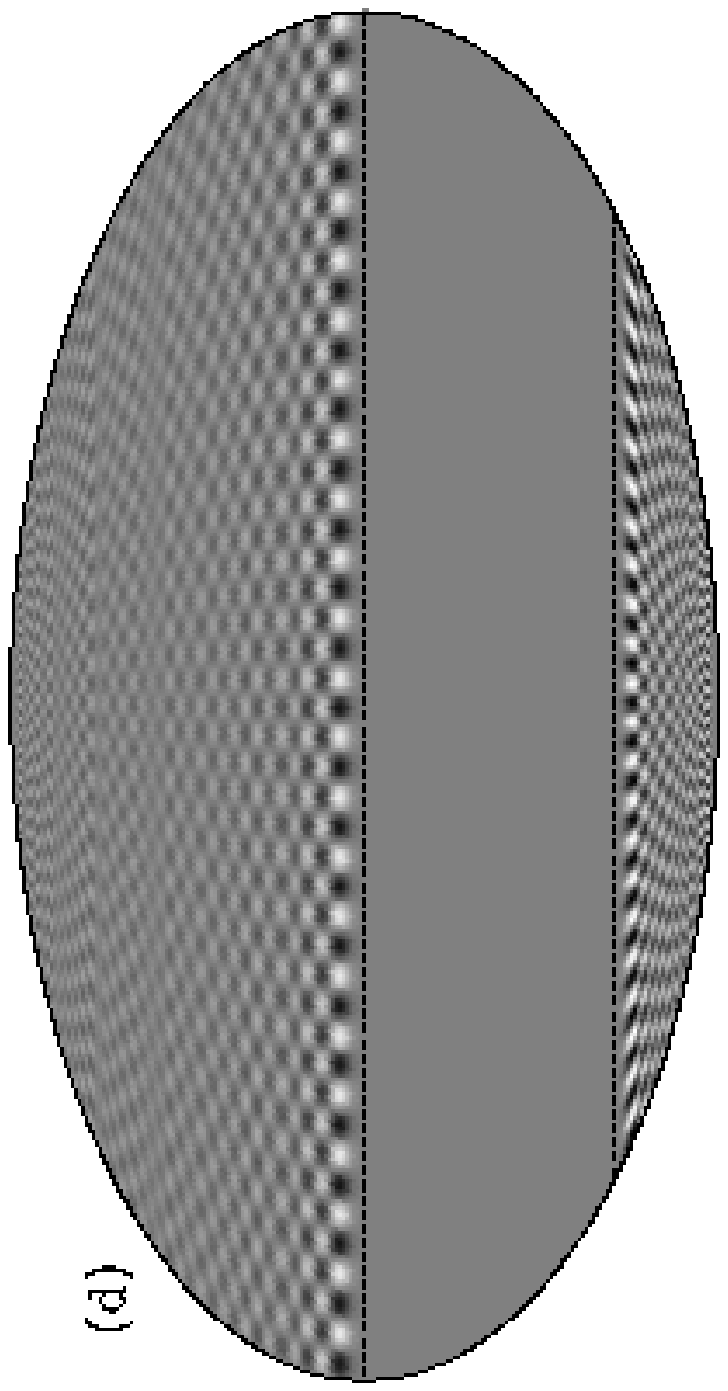}
\includegraphics{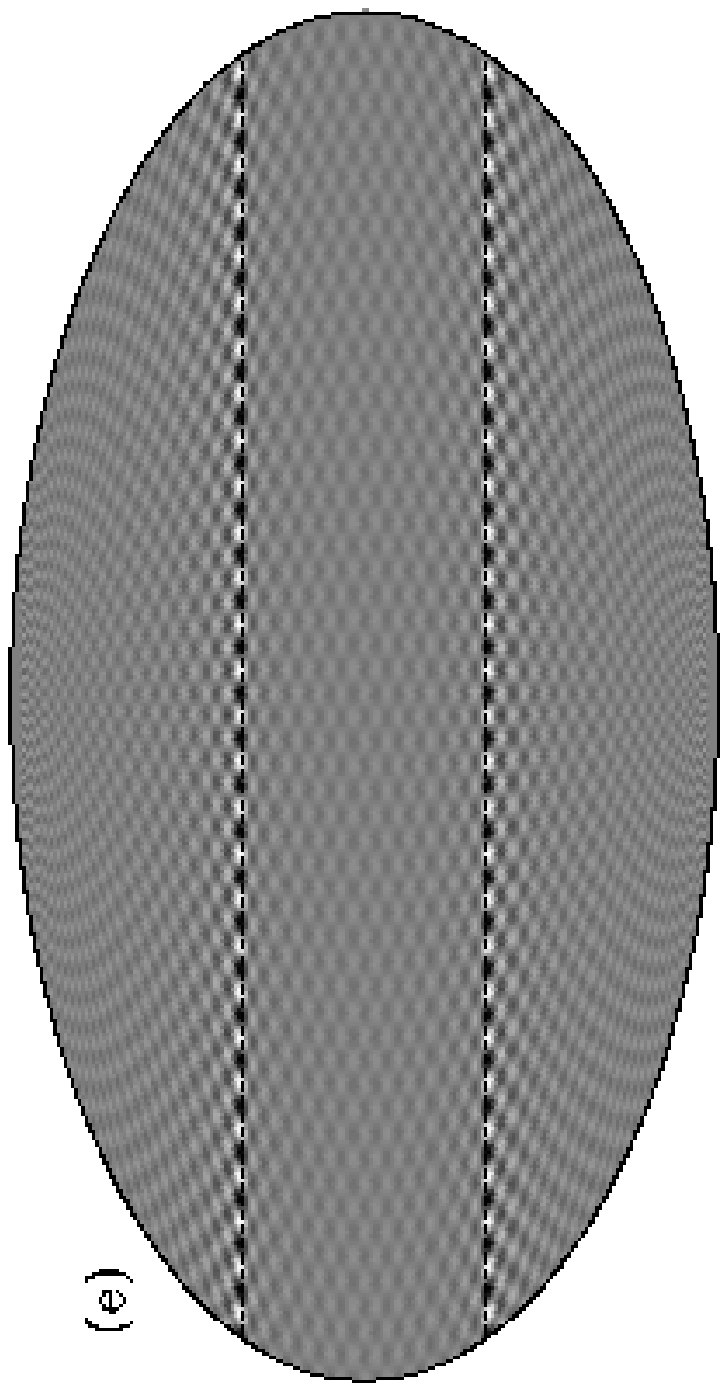}
\includegraphics{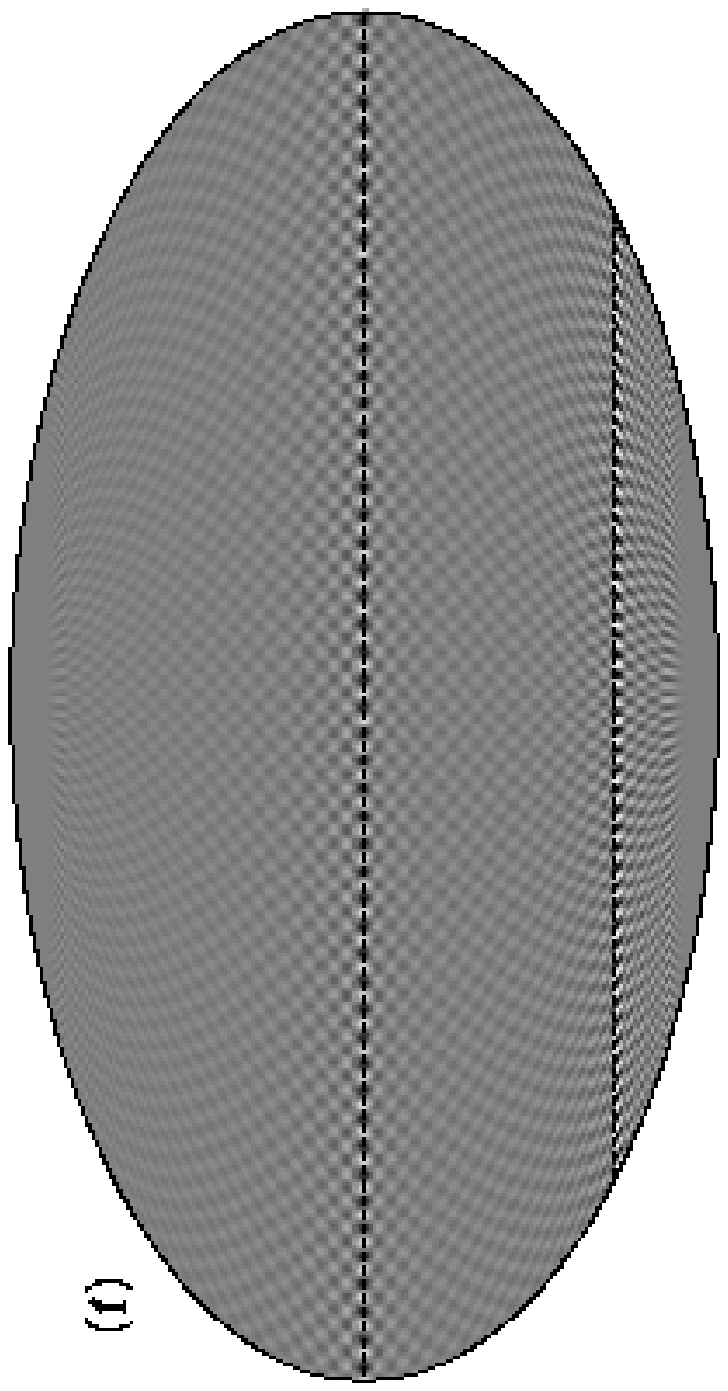}
\includegraphics{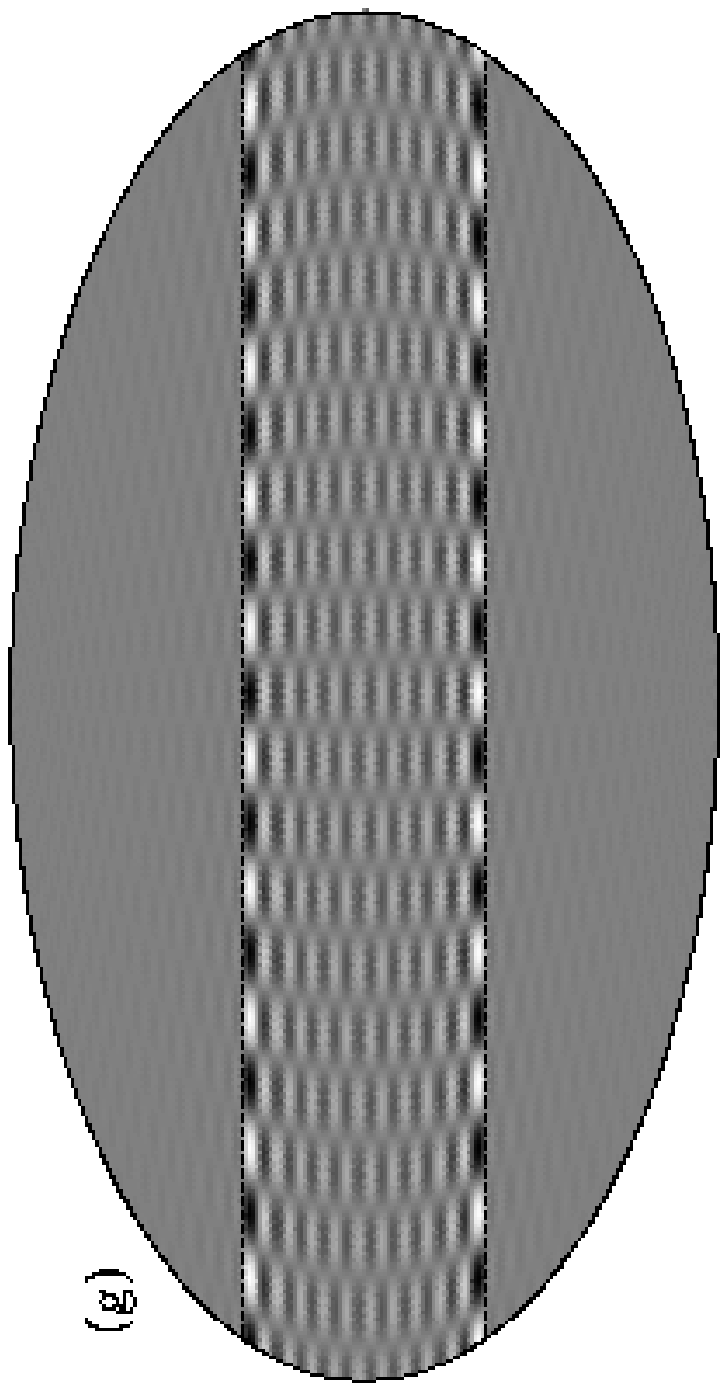}
\includegraphics{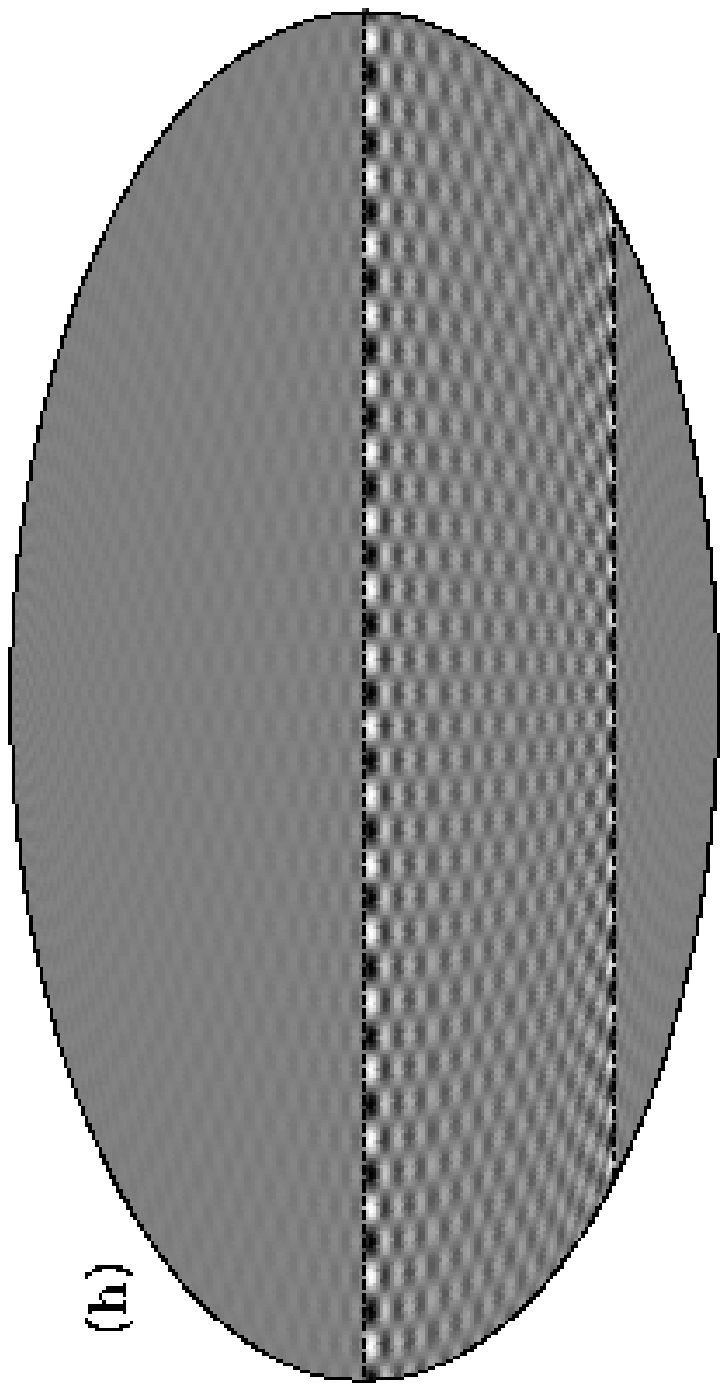}
\vspace{\triplefigureheight}
\caption{Orthonormal cut sphere basis functions,
$Y_i^\prime(\skyhat)$, as described in \sect{highres_orthog}.
In all cases 
$\lmax = 100$, $W_{\rmn min} = 0.01$, and a Mollweide projection is used.
The colour-map varies from black (large negative values), through
grey (zero), to white (large positive values), in each case being
scaled to cover the dynamic range of the relevant basis function.
For those in the left column [(a), (c), (e) and (g)] 
the cut (shown by the dashed lines)
is symmetric, with $b_{\rmn cut} = \pm 20$ deg;
for those in the right column [(b), (d), (f) and (h)]
the cut (again shown by the dashed lines)
is asymmetric, the region between 
$b_1 = 0$ deg and $b_2 = - 45$ deg having been removed.
Within each column the indexing of the basis functions 
is arbitrary, but they are displayed so that their fractional
support 
in the removed region increases
from (a) to (g) and (b) to (h), respectively.}
\label{figure:basis}
\end{figure*}

If $\lmax \la 50$ and most of the sphere is retained 
(\ie\ $\Omega_\cutsphere \ga \Omega_\sphere / 2 = 2 \pi$) then
the coupling matrix is numerically invertible and 
can be treated as 
positive definite in practice.
Consequently $\matr{A}$ 
[defined in \eq{coupling_decomp}]
is also invertible, and $\matr{B} = \matr{A}^{-1}$,
so that, from \eq{yprime_general},
the orthonormal basis set is given by 
\begin{equation}
\vect{Y}^\prime(\skyhat) = \matr{A}^{-1} \vect{Y} (\skyhat).
\end{equation}
The form of $\matr{A}$ depends on the factorization method;
of the wide variety available 
(\eg\ Golub \& van Loan 1996), 
the two most useful here are 
singular value decomposition (\svd)
and Cholesky decomposition.

The \svd\ of the covariance matrix 
is defined in terms of \eq{coupling_decomp} by
$\matr{A} = \matr{V} \matr{W}^{1/2}$
(\ie\ $\matr{C} = \matr{V} \matr{W} \matr{V}^{\rmn T}$),
where $\matr{V}$ is orthogonal and $\matr{W}$ is
diagonal\footnote{If $\matr{M}$ is
diagonal then the notation $\matr{M}^{\pm 1/2}$ is used
here to denote the matrix defined by
$(M^{\pm 1/2})_{i,j} = \delta_{i,j} M_{i,i}^{\pm 1/2}$,
where $\delta_{i,j}$ is the Kronecker delta function.
Thus $\matr{M}^{1/2}$ only exists if the diagonal elements of
$\matr{M}$ are non-negative
and $\matr{M}^{-1/2}$ only exists if the diagonal elements of
$\matr{M}$ are strictly positive.}.
The diagonal elements of $\matr{W}$ are the eigenvalues of 
$\matr{C}$ and, as their ordering is arbitrary,
$\matr{W}$ can be defined such that $W_{i,i} \leq W_{i+1, i+1}$,
provided the columns of $\matr{V}$ 
are permuted in the same way. 
The columns of $\matr{V}$, in turn, are the eigenvectors of $\matr{C}$, 
and $\matr{V}$ is an orthogonal
matrix (\ie\ $\matr{V}^{-1} = \matr{V}^{\rmn T}$).
Hence the conversion matrix is given by
$\matr{B} = \matr{A}^{-1} = \matr{W}^{-1/2} \matr{V}^{\rmn T}$,
which is trivially computed once the \svd\ has been performed. 
Note that this approach is effectively the same as the direct
calculation of the eigenstructure of $\matr{C}$ mentioned 
above in \sect{method}.

Whilst \svd\ is a powerful technique, it is computationally 
expensive -- a Cholesky decomposition is approximately ten
times faster, although it can only be performed on symmetric matrices
which are numerically positive definite. 
The Cholesky decomposition of the covariance matrix
takes the form $\matr{C} = \matr{L} \matr{L}^{\rmn T}$
(\ie\ $\matr{A} = \matr{L}$), where 
$\matr{L}$ is lower triangular. 
Hence the conversion matrix, 
$\matr{B} = \matr{L}^{-1}$ can be computed quickly from the 
initial factorization in this case as well.
The triangular structure of the conversion matrix also ensures 
that the new basis functions are the same as those formed 
by a numerically-stable Gram-Schmidt orthogonalisation 
(G\'{o}rski 1994).

Despite the fact that the \svd\ and the Cholesky decomposition 
result in quite different sets of basis functions, 
there is no reason to prefer one over the other in general.
In the case of \cmb\ analysis, however, the triangular structure
of $\matr{A}$ and $\matr{B}$ as generated by the Cholesky decomposition
is preferable as it ensures that the non-cosmological monopole 
and dipole modes are kept separate from the $l \geq 2$ modes,
assuming $l$-ordering is used (G\'{o}rski 1994).
If the \svd\ route is taken (or another indexing scheme used) the
separation of the $l = 0$ and $l = 1$ modes can be ensured using the
partial Householder transform described in \apdx{noncosmol}.
Nonetheless, if the coupling matrix is sufficiently non-singular,
a Cholesky decomposition should be used to create the orthonormal
basis set, due to both its computational efficiency and the simplicity
with which the non-cosmological modes are handled.

\subsubsection{High-resolution analysis}
\label{section:highres_orthog}

If $\lmax \ga 50$ the coupling matrix is numerically singular,
and thus $\matr{A}$ [defined in \eq{coupling_decomp}] is
non-invertible.
Cholesky decomposition of $\matr{C}$ is thus impractical
and, whilst an \svd\ is possible, the conversion
matrix as defined in \sect{lowres_orthog} cannot be computed, as 
the smallest elements of $\matr{W}$ (\ie\ the smallest 
eigenvalues of $\matr{C}$) are so close to zero. 
This implies that the corresponding columns of 
$\matr{V}$ do not contribute to the reconstruction of $\matr{C}$
and can be ignored. 
Hence it is possible to perform an approximate \svd\ of the coupling
matrix, defined by
$\matr{C} \simeq \tilde{\matr{V}} \tilde{\matr{W}} \tilde{\matr{V}}^{\rmn T}$
(\ie\ $\matr{A} = \tilde{\matr{V}} \tilde{\matr{W}}^{1/2}$),
where 
$\tilde{\matr{W}}$ is an 
$i^\prime_{\rmn max} \times i^\prime_{\rmn max}$ diagonal matrix 
containing the largest elements of $\matr{W}$ and 
$\tilde{\matr{V}}$ is 
an $i^\prime_{\rmn max} \times i_{\rmn max}$
matrix consisting of the corresponding columns of $\matr{V}$. 
The value of $i_{\rmn max}$ is determined by the choice of $W_{\rmn min}$
used to truncate $\matr{W}$, but the bimodality of the eigenvalue
distribution means that any value between $\sim 10^{-5}$ and $\sim 0.1$
is acceptable.
The resultant conversion matrix is $\matr{B} = \tilde{\matr{W}}^{-1/2}
\tilde{\matr{V}}^{\rmn T}$ 
(satisfying $\matr{B} \matr{A} \simeq \tilde{\matr{I}}$)
and the $i^\prime_{\rmn max}$ new 
basis functions are given by
\begin{equation}
\vect{Y}^\prime (\skyhat) = \tilde{\matr{W}}^{-1/2}
\tilde{\matr{V}}^{\rmn T} \vect{Y} (\skyhat).
\end{equation}

These basis functions represent an orthonormal basis set on the incomplete
sphere, but they are not formally complete to the nominal band-limit
due to the slightly approximate nature of the reduced \svd.
The decomposition becomes exact in the limit
$\lmax \rightarrow \infty$ as $\tilde{\matr{W}} \rightarrow \tilde{\matr{I}}$
(from the eigenvalue arguments described in \sect{eigen}) 
and the reduced \svd\ becomes
$\matr{C} \rightarrow \tilde{\matr{V}} \tilde{\matr{V}}^{\rmn T}$
(\ie\ $\matr{A} \rightarrow \tilde{\matr{V}}$ and
$\matr{B} \rightarrow \tilde{\matr{V}}^{\rmn T}$) in this limit.
Note also that these basis functions are still orthogonal on the full
sphere, although they are no longer normalised to unity; this 
is another potential advantage of the \svd-based method.

Several examples of orthonormal cut-sphere
basis functions are shown in \fig{basis},
for both symmetric and asymmetric constant latitude cuts. 
The link between these
functions and the spherical harmonics is apparent -- they have 
the same cellular structure, but, for the most part, are combined 
in such a way that their support is localised to $\cutsphere$. 
However the functions shown in \fig{basis} (g) and (h) are 
from the small subset with intermediate values of $W_{i,i}$
and, as such, have considerable support in the removed region. 

In the case of \cmb\ analysis, one
short-coming of this approach is that the non-cosmological monopole 
and dipole modes are not distinguished from the higher moments
(see \sect{highres_harm}),
although this separation can be achieved post facto 
by using a partial Householder transform, 
as described in \apdx{noncosmol}. 
In doing this some of the useful properties of the \svd\ are lost,
but this operation need only be performed as the last step in
generating the orthonormal basis set, by which stage all the 
computationally intensive matrix operations have already been performed.

Despite the inconvenience caused by the mixing of the monopole and dipole 
modes, \svd\ is clearly the most flexible and general
method of orthogonalisation.
In part, this stems from the fact that it can 
be applied to the coupling matrix without any prior knowledge 
of its singular properties. However, in the high-$\lmax$ limit,
the number of supported basis functions is determined by the area
of the cut and given, to a good approximation, by \eq{rank}.
Hence faster, if less powerful, techniques can be used to 
orthogonalise the spherical harmonics for $\lmax \ga 1000$ 
as the coupling matrix is guaranteed to have at 
least $(\lmax + 1)^2 \Omega_\cutsphere / (4 \pi)$ positive eigenvalues.
An example of this idea would be to 
modify the pivoting algorithm in the psuedo-Cholesky decomposition 
described in Section 4.2.9 of Golub \& van Loan (1996) so that 
the decomposition halts when the predetermined number of basis
functions have been generated, as opposed to using the less robust
threshold based on the values of the diagonal elements of $\matr{C}$. 

\subsection{Constant latitude cuts}
\label{section:constlat}

In principle the method presented above
is a complete solution to the problem
of constructing orthonormal bases on the cut sphere, but
the coupling matrix 
requires $O(\lmax^4)$ storage, 
limiting a general implementation to $\lmax \simeq 200$ 
on most current computers.
Furthermore, the \svd\ of an $n \times n$ matrix
requires $O(n^3)$ operations, and so the orthogonalisation
operation count scales as $O(\lmax^6)$.
Similar difficulties are encountered in merely evaluating $\matr{C}$,
regardless of whether numerical integration or recursive 
techniques are used (\eg\ Hivon \etal\ 2001).

Fortunately all these difficulties are significantly reduced
in the case of a constant latitude cut (\cf\ Oh, Spergel \& Hinshaw 1999;
Wandelt \etal\ 2001), 
defined by ignoring all $\theta$ for which
$\theta_1 \leq \theta \leq \theta_2$.
This could be the symmetric removal of the Galactic plane
(\ie\ $\theta_1 = \pi / 2 - b_{\rmn cut}$ and
$\theta_2 = \pi / 2 + b_{\rmn cut}$, where $b_{\rmn cut}$ is 
the latitude of the cut)
or the absence of data round one pole (\ie\ 
$\theta_1 = 0$ and $\theta_2 = \theta_{\rmn cut}$).
The formalism derived below can also be trivially extended to include
multiple cuts, as would be required for a \cmb\ experiment 
which did not observe either ecliptic pole.

Explicitly including the constant latitude cut in
\eq{coupling}, the elements of the coupling matrix are given by 
\begin{eqnarray}
\label{equation:covar_galcut}
C_{i(l, m), i(l^\prime, m^\prime)} & = &
\int_0^{2 \pi} s_{m}(\phi) s_{m^\prime}(\phi) \, {\rmn d} \phi
\\
& \times &
\left[
\int^{\cos(\theta_2)}_{-1} \lambda_{l, |m|}(x)
\lambda_{l^\prime, |m^\prime|}(x) \, {\rmn d}x 
\right.
\nonumber \\
& & + 
\left.
\int_{\cos(\theta_1)}^1 \lambda_{l, |m|}(x) 
\lambda_{l^\prime, |m^\prime|}(x) \, {\rmn d}x
\right], 
\nonumber
\end{eqnarray}
where $s_m(\phi)$ is defined in \eq{s_m},
and the
$\lambda_{l, m}(x)$ are normalised associated Legendre functions,
given in \eq{lambdalmdef}.
From \apdx{sphar}, the first integral in \eq{covar_galcut} 
reduces to 
$2 \pi \delta_{m, m^\prime}$,
and so
\begin{eqnarray}
\label{equation:covar_simple}
C_{i(l, m), i(l^\prime, m^\prime)} 
& \!\!\!\! = \!\!\!\! &
\delta_{m, m^\prime} \\
& \!\!\!\! \times \!\!\!\! &
\left[\delta_{l, l^\prime} - 
2 \pi \int_{\cos(\theta_2)}^{\cos({\theta_1})} \lambda_{l, |m|}(x)
\lambda_{l^\prime, |m|}(x) \, {\rmn d}x \right]
\!\!. 
\nonumber
\end{eqnarray}
The remaining integral can be evaluated using a combination of 
analytical formul\ae\ and recursion relations, as described in
\apdx{integ}.

The most important aspect of \eq{covar_simple}
is that the coupling matrix $\matr{C}$ is extremely sparse
(only one element in $\sim l_{\rmn max}$ is non-zero) and,
if stored using the indexing scheme defined in \eq{indexing}
(\ie\ grouped into sub-matrices of fixed $m$), is block 
diagonal. 
$\matr{C}$ can thus be stored in the form
of $2 l_{\rmn max} + 1$ sub-matrices, the
$m$th of which has 
$(l_{\rmn max} + 1 - |m|)^2$ elements,
and the storage requirements thus scale as 
$O(\lmax^3)$ rather than $O(\lmax^4)$.
Whilst it is convenient to store all the blocks 
simultaneously, there is no need to do so,
which can further reduce the storage requirements to $O(\lmax^2)$.
It is also clear from \eq{covar_simple}
that only the $m \geq 0$ terms need be treated explicitly
and that $l$ and $l^\prime$ are interchangeable, decreasing the 
storage requirements by a further factor of four.
Finally, in the case of a symmetric cut (\ie\ $\theta_2 = \pi - \theta_1$)
the parity of $\lambda_{l,m}(x)$ is such that all terms for which
$l + l^\prime$ is odd vanish, resulting in an additional halving 
of the memory requirements.

The orthogonalisation can be performed by 
decomposing each sub-matrix 
separately, reducing the operation count from
$\order(l_{\rmn max}^6)$ to $\order(l_{\rmn max}^4)$.
The removal of the poorly-supported basis functions is achieved 
in the same manner as described in \sect{method}, although
the book-keeping is more complicated.
Similarly the partial Householder transform required to 
separate the $l = 0$ and $l = 1$ modes 
need only be applied 
to the $m = 0$ and $m = \pm 1$ blocks of the resultant conversion
matrix (\apdx{noncosmol}).
An important side-effect of the separation in $m$ 
is that the $Y^\prime_i(\skyhat)$ have the same 
trigonometric $\phi$-dependence as the full-sky spherical harmonics 
(\apdx{sphar}). This also implies that the $Y^\prime_i(\skyhat)$
can be treated as two-index quantities, defined by $m$ and 
a second, arbitrary index in place of $l$.

The algorithms described here were implemented 
on the Cambridge 
Centre for Mathematical Science's
COSMOS 64-processor Silicon Graphics Origin 2000 and
the evaluation and decomposition of the coupling matrix
at the highest \Planck\ resolution of $\lmax \simeq$ 2500
required about an hour.
The majority of the time was spent factorizing
the sub-matrices,
and thus significant accelerations are unlikely,
the highly-optimised 
{\sc linear algebra package} (\lapack; Anderson 1992)
routines having been used for all the decompositions.
For a given choice of $\theta_1$ and $\theta_2$,
the decomposition of $\matr{C}$ need only be performed once, 
so orthogonalisation of the spherical harmonics on an incomplete 
sky should comprise only a small fraction of the analysis
required for the forthcoming \MAP\ and \Planck\ missions.

\section{Harmonic analysis}
\label{section:harmonics}

Methods for constructing an orthonormal
basis set on the incomplete sphere from the spherical harmonics
were discussed in \sect{orthog},
but in most cases of data analysis it is the harmonic coefficients,
representing functions on the sphere, that are of interest. 
There are at least three useful harmonic expansions of a general
function on the sphere, and the relationships between these coefficients,
which are summarised in \tabl{harm}, are derived here.

\begin{table*}
\begin{minipage}{117mm}
\caption{Conversions between harmonic coefficients}
\begin{tabular}{llll}
\hline
& & & \\
$\vect{a} = \int_\sphere \vect{Y}(\skyhat) a(\skyhat) \, {\rmn d}\Omega$ &
        & $\hat{\vect{a}} = \matr{B}^{\rmn T} \matr{B} \tilde{\vect{a}}
                \rightarrow \matr{C}^{-1} \tilde{\vect{a}}$
        & $\hat{\vect{a}} = \matr{B}^{\rmn T} \vect{a}^\prime
                \rightarrow 
		\left(\matr{A}^{\rmn T}\right)^{-1} \vect{a}^\prime$ \\
& & & \\
$\tilde{\vect{a}} = \int_\cutsphere \vect{Y}(\skyhat) 
                a(\skyhat) \, {\rmn d}\Omega $ &
        $\tilde{\vect{a}} = \matr{C} \vect{a}$
        &
        & $\tilde{\vect{a}} = \matr{A} \vect{a}^\prime$ \\
& & & \\
$\vect{a}^\prime = \int_\cutsphere \vect{Y}^\prime(\skyhat) 
                a(\skyhat)\,{\rmn d}\Omega $ &
        $\vect{a}^\prime = \matr{A}^{\rmn T} \vect{a}$
        & $\vect{a}^\prime = \matr{B} \tilde{\vect{a}}$
        & \\
& & & \\
\hline
\label{table:harm}
\end{tabular}

The conversions between the various harmonic coefficients defined
in \sect{harmonics}:
$\vect{a}$ are the standard coefficients of the spherical harmonics
[\eq{a_def}];
$\tilde{\vect{a}}$ are the psuedo-harmonic coefficients [\eq{apsuedo_def}];
$\vect{a}^\prime$ are the cut-sphere harmonic coefficients [\eq{aprime_def}];
and
$\hat{\vect{a}}$ are the reconstructed spherical harmonic coefficients.
The two basis functions are the spherical harmonics, $\vect{Y}(\skyhat)$ 
	[defined in \apdx{sphar}],
and the orthogonalised harmonics, $\vect{Y}^\prime(\skyhat)$
	[defined in \sect{orthog}].
If the coupling matrix of the spherical harmonics, $\matr{C}$, is
invertible, then the expressions for $\hat{\vect{a}}$ following the
$\rightarrow$ can be used as exact inversions; otherwise the `estimators'
are approximate projections onto the cut region.
$\matr{A}$ can be any matrix such that
$\matr{A} \matr{A}^{\rmn T} = \matr{C}$,
and
$\matr{B}$ can be any matrix such that
$\matr{B} \matr{A} = \tilde{\matr{I}}$.

\end{minipage}
\end{table*}

A band-limited function, $a(\skyhat)$, can be completely specified
by a finite number of harmonic coefficients as (\cf\ \apdx{sphar})
\begin{equation}
\label{equation:f_a}
a(\skyhat) = \vect{Y}^{\rmn T}(\skyhat) \vect{a},
\end{equation}
where it is assumed that $\lmax$ is greater than or equal to the 
band-limit of $a(\skyhat)$ and 
the harmonic coefficients are defined by
\begin{equation}
\label{equation:a_def}
\vect{a} = \int_\sphere \vect{Y}(\skyhat) a(\skyhat) \, {\rmn d}\Omega .
\end{equation}
The invertibility of these transformations is due to the 
orthonormality of the spherical harmonics on $\sphere$ 
and the fact that they represent a complete basis set given
the band-limit.

If $a(\skyhat)$ is only known over some fraction of the sphere
$\cutsphere \leq \sphere$, 
then $\vect{a}$ cannot be determined as above, 
as the integral in \eq{a_def} is incomplete.
In this case the psuedo-harmonics 
\begin{equation}
\label{equation:apsuedo_def}
\tilde{\vect{a}} = \int_\cutsphere \vect{Y}(\skyhat) a(\skyhat) 
\, {\rmn d}\Omega 
\end{equation}
fully specify $a(\skyhat)$ in $\cutsphere$ due to the band-limit. 
From \eq{f_a} they are related to the full harmonic coefficients by
\begin{equation}
\label{equation:apsuedo_a}
\tilde{\vect{a}} = \matr{C} \vect{a},
\end{equation}
where $\matr{C}$ is the coupling matrix, defined in \eq{coupling}.

The psuedo-harmonics are useful quantities, but it is preferable to
work with basis functions that are orthonormal on 
$\cutsphere$.
Denoted $\vect{Y}^\prime (\skyhat)$ in \sect{orthog}, their 
harmonic coefficients are given by
\begin{equation}
\label{equation:aprime_def}
\vect{a}^\prime
= \int_\cutsphere \vect{Y}^\prime(\skyhat) a(\skyhat) \, {\rmn d}\Omega .
\end{equation}
The relationship between $\vect{Y}(\skyhat)$ and 
$\vect{Y}^\prime(\skyhat)$ given in \eq{yprime_general}
flows through to the harmonic 
coefficients and applying
\eqs{yprime_general}, 
(\ref{equation:coupling_decomp}) and 
(\ref{equation:f_a})
to \eq{aprime_def} gives
\begin{eqnarray}
\label{equation:aprime_a}
\vect{a}^\prime 
& = & \matr{B} \int_{\cutsphere}
\vect{Y}(\skyhat)
\left[
\vect{Y}^{\rmn T}(\skyhat) \vect{a}
\right]
\, {\rmn d} \Omega \nonumber \\
& = & \matr{B} \matr{C} \vect{a} \nonumber \\
& = & \matr{A}^{\rmn T} \vect{a} .
\end{eqnarray}

\begin{figure*}
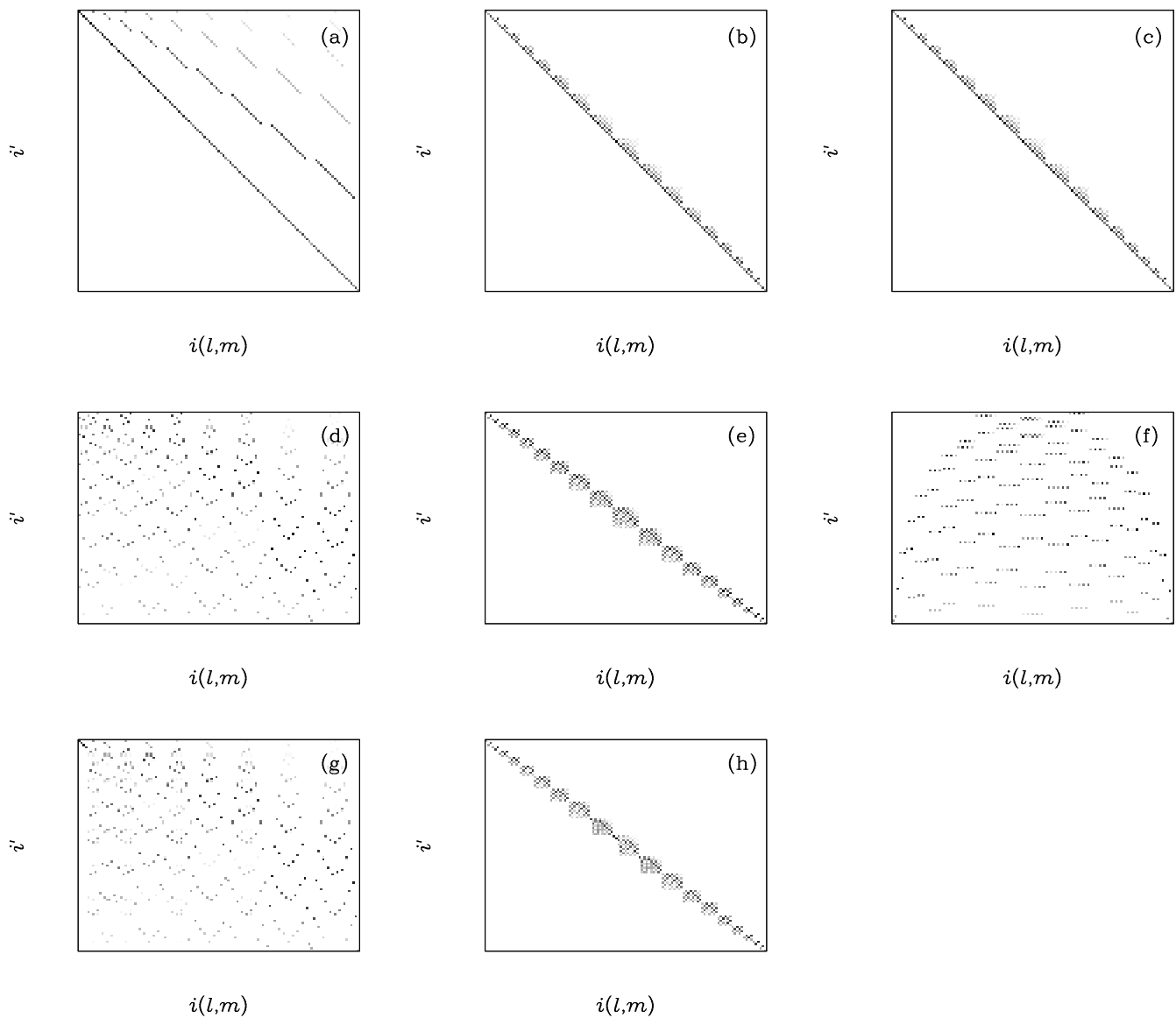

\includegraphics{clchol.ps}
\includegraphics{cmbchol.ps}
\includegraphics{cmchol.ps}
\includegraphics{clsvd.ps}
\includegraphics{cmbsvd.ps}
\includegraphics{cmsvd.ps}
\includegraphics{clhh.ps}
\includegraphics{cmbhh.ps}
\vspace{16.0cm}
\caption{Several examples of the conversion matrices, $\matr{A}^{\rmn T}$,
that relate
the orthonormal cut sphere harmonic coefficients, $\vect{a}^\prime$
to the conventional spherical harmonic coefficients, $\vect{a}$,
according to \eq{aprime_a}.
In all cases a symmetric, constant latitude of $b_{\rmn cut} = \pm 20$ deg
has been applied and $\lmax = 10$. 
The row index, $i^\prime$, corresponds to the cut-sky harmonic coefficients 
and 
the column index, $i(l,m)$, corresponds to the original spherical 
harmonic coefficients.
The colour-map shows the absolute value of the elements of $\matr{A}^{\rmn T}$
varying from zero (white) to their maximum value (black), 
which is normalised separately in each case.
The spherical harmonic coefficients are indexed
using $l$-ordering
in the left column [(a), (d) and (g)];
$m$-ordering has been used in order to utilise
the decoupling of different $m$-modes resulting from the azimuthal symmetry
by using block-by-block storage
in the central column [(b), (e) and (h)];
and 
$m$-ordering is used, but the structure of the coupling
matrix is not utilised, in the right column
[(c) and (f)].
In the case of $l$-ordering the non-cosmological modes are the four
left-most columns of the conversion matrix;
in the case of $m$-ordering the non-cosmological modes are the 
left-most columns of the three central blocks.
The conversion matrices shown in the top row
[(a), (b) and (c)] result from a Cholesky decomposition;
those in the middle row [(d), (e) and (f)] 
are produced by an \svd\ (in which all modes
with $W_{i,i} \leq 0.1$ have been removed);
finally, in the bottom row [(g) and (h)], 
the partial Householder transform 
described in \apdx{noncosmol} is applied to the conversion matrices 
shown in the above panels [(d) and (e), respectively].}
\label{figure:conversion}
\end{figure*}

The form of the above transformation depends on the decomposition
used to generate $\matr{A}$ (\cf\ \sect{method}), but, for \cmb\
analysis, it is desirable to separate the non-cosmological modes.
This amounts to demanding that the only four of the $a^\prime_i$
have any contribution from the $l = 0$ and $l = 1$ spherical harmonic
coefficients. 
The uppper triangular structure of $\matr{A}^{\rmn T} = \matr{L}$ as
generated by a Cholesky decomposition inherently satisfies this requirement,
but in general the conversion matrix needs to be transformed 
explicitly. One option is to use successive partial Householder
transforms, as described in detail in \apdx{noncosmol}. 
As can be seen from \fig{conversion}, 
the index-ordering and decomposition method 
combine to give a wide variety
of conversion matrices; which of these is most suitable depends on 
the application.

It is also possible to convert between the psuedo-harmonics and
the cut-sky harmonics, as they both contain information about
$a(\skyhat)$ in $\cutsphere$ alone.
Combining 
\eqs{coupling_decomp},
(\ref{equation:apsuedo_def})
and 
(\ref{equation:aprime_def}) implies that
$\tilde{\vect{a}} = \matr{A} \vect{a}^\prime$
and 
$\vect{a}^\prime = \matr{B} \tilde{\vect{a}}$.
However it is not always possible to determine
$\vect{a}$ from either $\vect{a}^\prime$ or
$\tilde{\vect{a}}$. 
In the low-$\lmax$ limit these inversions are defined
(\sect{lowres_harm}), but for appreciable band-limits 
only a projection onto the cut sphere is possible
(\sect{highres_harm}). 

\subsection{Low-resolution analysis}
\label{section:lowres_harm}

If the coupling matrix is numerically non-singular 
(\ie\ $\lmax \la 50$) then 
\eq{aprime_a} can be inverted to give
\begin{equation}
\label{equation:a_aprime}
\vect{a} = {\left(\matr{A}^{\rmn T}\right)}^{-1} \vect{a}^\prime
= \matr{B}^{\rmn T} \vect{a}^\prime,
\end{equation}
and \eq{apsuedo_def} implies that
\begin{equation}
\vect{a} = \matr{C}^{-1} \tilde{\vect{a}}.
\end{equation}
These are specific examples of the fact that
a band-limited function is completely defined
if it is known over any finite portion of the sphere,
and a cut-sky analysis 
serves no purpose --
any apparently localised contaminants infect
the entire sky. 
However any measurement of a field on the sphere is 
subject to noise which is not band-limited, in which case the 
application of a cut has the effect of greatly amplifying the
noise in the removed region (see \sect{powspec}), justifying
the use of a cut-sky analysis in the low-resolution limit. 

\subsection{High-resolution analysis}
\label{section:highres_harm}

If $\lmax \ga 50$ and the coupling matrix is numerically
singular, it is impossible to reconstruct (even) a band-limited
function that is known only on $\cutsphere$. 
The loss of information about modes constrained to the cut 
makes it clear that the analysis has the desired effect 
of removing contaminated (or otherwise problematic) regions, 
but the most appropriate transformation from the cut-sphere
basis to conventional harmonics is less obvious.

A least squares-like approach leads to a definition
of the reconstructed full sky coefficients as
\begin{equation} 
\label{equation:a_inf}
\hat{\vect{a}} 
\simeq 
\matr{B}^{\rmn T} \vect{a}^\prime
= 
\left(\matr{A} \matr{B} \right)^{\rmn T} \vect{a}.
\end{equation}
Similarly, \eq{apsuedo_a} implies that
\begin{equation}
\hat{\vect{a}}
\simeq
\matr{B}^{\rmn T} \matr{B} \tilde{\vect{a}}
=
\left(\matr{A} \matr{B} \right)^{\rmn T} \vect{a},
\end{equation}
where 
$(\matr{A} \matr{B})^{\rmn T}$ is a projection 
operator\footnote{The 
definition of $\matr{B}$ given in \sect{method} implies that 
$[{(\matr{A} \matr{B})^{\rmn T}}]^2 = (\matr{A} \matr{B})^{\rmn T}$
and it is hence a projection operator if $\lmax \rightarrow \infty$.}
onto the range of $\matr{C}$, which in real space is $\cutsphere$.
If $\lmax \rightarrow \infty$ it is possible to write
$\hat{a}(\skyhat) = w_{\cutsphere}(\skyhat) a(\skyhat)$,
where $w_{\cutsphere}(\skyhat)$ is the sharp window
function defined in \eq{window}. 

It is at this point that the subtle distinctions between the use of a 
discrete cut and a window function become apparent. 
These results only hold for band-limited functions, but, 
as defined above,
$\hat{a}(\skyhat)$ is not band-limited, and so cannot be analysed
self-consistently. Whether a discrete cut or 
an apodizing function is to be preferred
 depends on the situation in which the
incomplete sky analysis is required.

\section{\cmb\ data analysis}
\label{section:anal}

In order to determine the properties of the \cmb\ from noisy
observations of the microwave sky (\sect{obs}) a number of 
non-trivial analysis steps are required, including: map-making 
(\sect{map}); component separation (\sect{separate})
and
power spectrum estimation (\sect{powspec}). 
Several algorithms have been suggested for all these steps,
and these are discussed briefly below, but the main focus 
is on when and how to apply a sky cut. 
Further, whilst it is possible to analyse the data in either real
or Fourier space,
the latter approach is emphasised here as it
is more directly related to the formalism described in \sects{orthog}
and \ref{section:harmonics}, as well as being the focus 
of a related series of papers (van Leeuwen \etal\ 2001;
Challinor \etal\ 2001; Stolyarov \etal\ 2001). 
Note that the term `map' is used here to denote any representation of a 
field on the sky and can imply either a set of real space pixel
values or a vector of spherical harmonic coefficients.

\subsection{Observations}
\label{section:obs}

Observations of the \cmb\ can be made using a number of quite distinct
techniques.
Data have been obtained from the ground, high altitude balloons, 
and satellites, but the more important distinction is the type of 
telescope.
The experiments listed in \sect{intro} include:
straightforward single dish telescopes,
such as \Boomerang\ (\eg\ Netterfield \etal\ 2001)
and \Planck\ (\eg\ Bersanelli \etal\ 1996);
differencing experiments,
such as \COBE\ (\eg\ Smoot \etal\ 1992)
and \MAP\ (\eg\ Jarosik \etal\ 1998);
and interferometers, such as
the Cambridge Anisotropy Telescope (\CAT; Scott \etal\ 1996),
the Cosmic Backround Imager (\CBI; Padin \etal\ 2001)
and 
the Degree Angular Scale Interferometer (\DASI; Halverson \etal\ 2001).
The interferometry surveys inevitably cover only a small 
fraction of the sky, and so a flat-sky Fourier analysis becomes 
possible.
However the both the differencing and single dish surveys can,
in principle, cover most of the celestial sphere, and should yield 
maps of the microwave sky that are limited only by (the combined effects 
of) instrument noise and the finite telescope beam. 

\subsubsection{Noise}
\label{section:noise}

The typical receivers used in the above experiments have two main 
noise contributions: random white noise
and a correlated low-frequency (\ie\ `$1/f$') component.
The latter is potentially troublesome, leading to `stripy' maps
with correlated errors, and is the main reason for the popularity 
of differencing experiments which remove low frequency noise at the 
moment of observation. However data from single dish surveys 
can be `de-striped' if the scan strategy includes sufficiently
many multiply-observed points (\eg\ Tegmark 1997;
Delabrouille 1998; Maino \etal\ 1999) or the time-time noise covariance
matrix can be fully included in the map-making process
(Wright, Hinshaw \& Bennett 1996; Natoli \etal\ 2001; Challinor \etal\ 2001). 
Hence correlated errors are ignored in the simple analysis
presented here.

This leaves only the white component,
which can be analysed most simply in the case of a single beam experiment.
Following Knox (1995), a
receiver is characterised by its sensitivity,
$s$ (generally chosen to have
units of temperature time$^{1/2}$).
Assuming the noise is Gaussian it has 
expectation values 
$\langle n \rangle = 0$ and $\langle n^2 \rangle = s^2 t$ 
over an integration time $t$. 
The manner in which this noise projects onto a sky map depends on the
map-making algorithm, the scan strategy, and the beam.

\subsubsection{Beam convolution}
\label{section:convolve}

All telescopes necessarily have a finite point-spread function or 
beam, which, for a given detector,
can be characterised by $b(\skyhat)$, the fraction 
of photons from direction $\skyhat$ that are registered,
given a nominal orientation towards the north pole (\ie\ $\theta = 0$).
The harmonic expansion of the beam in this orientation is denoted
$\vect{b} = b_{i(l,m)}$, with 
the band-limit being related to the nominal
resolution of the detector.
For a given type of telescope the resolution 
improves with frequency due to diffraction effects;
this places limitations on the component separation algorithms 
that are used on the incomplete sky (\sect{separate}).

Most experiments have beams that are manifestly asymmetric, 
a fact which must be accounted for explicitly by the data
analysis algorithms, but the cut-sky issues of interest here 
can be explored more clearly if the beam is approximated by its
azimuthally averaged counterpart (\eg\ Challinor \etal\ 2001).
Defined by 
\begin{equation}
\bar{b}(\theta) =
\frac{1}{2 \pi} \int_0^{2 \pi} b(\theta, \phi) \, {\rmn d}\phi,
\end{equation}
its harmonic coefficients are simply
\begin{eqnarray}
\bar{b}_{l,m} & = & \delta_{m,0} b_{l,0} \nonumber \\
& = & 
\delta_{m,0} 
\sqrt{(2 l + 1) \pi}
\int_{-1}^{1} \bar{b}\left[\arccos(x)\right] P_l(x) \, {\rmn d}x ,
\end{eqnarray}
where $P_l(x)$ is a Legendre polynomial (\apdx{sphar}).
The use of $\bar{b} (\skyhat)$ allows the definition of a 
beam-smoothed sky, $\smooth{s}(\skyhat)$, 
given in terms of the true sky, $s(\skyhat)$, by
\begin{equation}
\label{equation:s_smooth}
\smooth{s}(\skyhat) = \int_{\sphere}
\bar{b} \left[\arccos \left(\skyhat \cdot \skyhat^\prime \right) \right]
s(\skyhat^\prime) \, {\rmn d} \Omega^\prime.
\end{equation}
This convolution is much simpler in harmonic space, and applying 
\eq{a_lm} to \eq{s_smooth} yields 
\begin{equation}
\label{equation:convolve_fullsky}
\smooth{\vect{s}} = \bar{\matr{B}} \vect{s},
\end{equation}
where $\bar{\matr{B}}$ (as distinct from the conversion matrix, 
$\matr{B}$) is a diagonal `convolution matrix' with 
\begin{eqnarray}
\bar{B}_{i,i} & = & \sqrt{\frac{4 \pi}{2 l(i) + 1}}
\bar{b}_{l(i), 0} \nonumber \\
& = & 
2 \pi
\int_{-1}^{1} \bar{b}\left[\arccos(x)\right] P_{l(i)}(x) \, {\rmn d}x ,
\end{eqnarray}
$l(i)$ 
being defined in \apdx{sphar}.
The simple form of \eq{convolve_fullsky} is often utilised 
explicitly in \cmb\ analysis algorithms 
(\eg\ Knox 1995; Hobson \etal\ 1998; Oh \etal\ 1999; Stolyarov 2001),
but in all cases full sky coverage is -- sometimes implicitly -- assumed. 

Turning to convolution on the incomplete sphere, $\cutsphere$, 
application of \eq{aprime_a} to \eq{convolve_fullsky} yields
\begin{equation}
\smooth{\vect{s}}^\prime = 
\matr{A}^{\rmn T} \bar{\matr{B}} \vect{s},
\end{equation}
where $\matr{A}$ is defined implicitly in \eq{coupling_decomp}.
In the low-resolution limit \eq{a_aprime} then gives 
the cut-sky analogue of \eq{convolve_fullsky} as 
\begin{equation}
\label{equation:convolve_cutsky}
\smooth{\vect{s}}^\prime = 
\bar{\matr{B}}^\prime
\vect{s}^\prime,
\end{equation}
with the new convolution matrix defined by 
$\bar{\matr{B}}^\prime = 
\matr{A}^{\rmn T} \bar{\matr{B}} \matr{B}^{\rmn T}$. 

For higher band-limits no such relation exists,
the loss of 
modes in the cut region rendering the convolution ill-defined. 
This is simply understood in real space, as the value of $\smooth{s}
(\skyhat)$ near the edge of $\cutsphere$ is given by an integral that
extends several beam widths into the removed region. 
Thus it is impossible to relate $\smooth{s} (\skyhat)$ to 
$s (\skyhat)$ with $\skyhat \in \cutsphere$. 
These arguments are true independent
of the representation chosen, but in harmonic space they mean that it is
impossible to relate $\smooth{\vect{s}}^\prime$ to 
$\vect{s}^\prime$. 

Whilst \eq{convolve_cutsky} is formally incorrect in high-$\lmax$ 
cases, it is potentially useful as a practical approximation. 
It
is equivalent to assuming that the signal 
is given by $\vect{Y}^\prime(\skyhat) \vect{s}^\prime$,
which implies that $s(\skyhat) \simeq 0$ in $\sphere - \cutsphere$.
This 
is particularly inaccurate if the 
removed region contains
anomalously strong sources, such as the Galactic plane.
Nonetheless, \eq{convolve_cutsky} 
gives $s^\prime(\skyhat)$ correctly for all 
$\skyhat$ more than a few beam widths away from the 
edge of the cut region. 
However,
even if this is an acceptable approximation, there is the further
inconvenience that the effective cut-sky beam, $\bar{\matr{B}}^\prime$,
is not diagonal, introducing couplings between all the modes.

In short, it is preferable to avoid performing any sort of convolution
(or deconvolution) on the cut sky, although it is clear
that this situation is encountered in any survey with incomplete
sky coverage. The one, albeit trivial,
exception to this rule is if the beam is a delta function, 
or at least the closest approximation to a delta function 
possible given the band-limit under consideration.
In this case
$\bar{\matr{B}} = \matr{I}$ and hence, 
from \eq{b_decomp}, $\bar{\matr{B}}^\prime = \tilde{\matr{I}}$ as well.
\Eq{convolve_cutsky} then implies that 
$\smooth{\vect{s}}^\prime$ ($= \vect{s}^\prime$) is
the true, unconvolved sky map, estimation of which is addressed 
next. 

\subsection{Map-making}
\label{section:map}

Some of the most important products of the next generation 
of \cmb\ survey will be 
high-resolution maps of the sky at each of several frequencies. 
Such maps can be created in a number of ways, but care must be 
taken to account for a huge variety of systematics whilst retaining as
much information as possible. 
Both real space
(\eg\ Wright \etal\ 1996; Bond \etal\ 1998; Natoli \etal\ 2001)
and Fourier space
(van Leeuwen \etal\ 2001; Challinor \etal\ 2001)
algorithms have been proposed as being suited to particular 
apsects of the map-making problem. 
The resultant uncertainties 
in the pixel values or harmonic coefficients
depend on both the data itself (\ie\ the scan strategy,
noise properties, \etc) and the 
map-making algorithms used, and can vary quite markedly
from experiment to experiment. 

Here only the idealised case of uniform sky coverage is considered 
as the discussion which follows is not significantly changed by this
useful simplification. 
Under this assumption the optimal
estimator for the unsmoothed sky, $\hat{\vect{s}}$ ($= \vect{s} + \vect{n}$),
would be unbiased 
(\ie\ $\langle \hat{\vect{s}} \rangle = \vect{s}$)
and have covariance given by 
\begin{equation}
\label{equation:covariance}
\matr{N} =
\left\langle \vect{n} \vect{n}^{\rmn T} \right\rangle
=
\left\langle 
\left( \hat{\vect{s}} - \vect{s} \right)
\left( \hat{\vect{s}} - \vect{s} \right)^{\rmn T}
\right\rangle
= \sigma^2 \bar{\matr{B}}^{-2},
\end{equation}
where (\cf\ Knox 1995)
$\sigma^2 = 4 \pi s^2 / (N_{\rmn d} t_{\rmn obs})$,
$t_{\rmn obs}$ is the total observation time of the survey,
and 
$N_{\rmn d}$ is the number of detectors at the frequency in question
(all of which are assumed to have the same beam).
An important issue at this point is the band-limit chosen.
Clearly only a finite analysis is possible in practice, and 
$\bar{\matr{B}}$ becomes increasingly singular as $\lmax \rightarrow
\infty$; these two points are related in so far as the sky can
never be reconstructed with infinite resolution. 
The choice of $\lmax$ is somewhat arbitrary, although any value a factor
of a few greater than the effective beam width will ensure that
$\bar{\matr{B}}^{-1}$ exists whilst discarding only multipoles 
that are noise-dominated. 
The fact that, unlike the useful signal, the noise is not subject
to any band-limit is critical to the understanding of the low-$\lmax$
cut-sky power spectrum estimation discussed in \sect{powspec}.

Note also that, due to the assumption of uniform sky coverage, 
the covariance matrix is diagonal. Transforming this estimator
into real space yields maps with covariance given by
\begin{equation}
\label{equation:covariance_real}
\langle n(\skyhat_1) n(\skyhat_2) \rangle
= 
\sigma^2
\vect{Y}^{\rmn T} (\skyhat_1) \bar{\matr{B}}^{-2}
\vect{Y} (\skyhat_2).
\end{equation}
As the noise term in the data is not beam-convolved the removal of
the beam results in spatial correlations of the noise (as encoded 
in $\bar{\matr{B}}$), as well as correlations due to the 
finite resolution analysis (the sums over spherical harmonics), 
which are essentially equivalent to pixel smoothing.
In the more realistic case of non-uniform sky coverage, the covariance
matrix is non-diagonal in both bases, a point discussed further
by Oh \etal\ (1999).

The above estimator for the true sky is closely linked 
to the more commonly used estimator for the
smoothed sky, $\hat{\smooth{\vect{s}}}$. 
Being related by
$\hat{\smooth{\vect{s}}} = \bar{\matr{B}}^{-1} 
\hat{\vect{s}}$, it is clear they contain the same information
(under the assumption of a symmetric beam).
The covariance structure of $\hat{\smooth{\vect{s}}}$ is simpler 
as the correlations discussed above are not introduced,
but $\hat{\vect{s}}$ is a more natural data object 
in the context of this discussion as it is the true
sky that is of interest.
In particular, unsmoothed maps allow more flexibility in applying 
a sky cut, as the problems with convolution on the incomplete 
sphere described in \sect{convolve} do not arise.
In practice the best compromise may be to reconstruct the 
sky convolved with the azimuthally averaged beam, thus 
creating maps with the simplest covariance structure possible
without information loss. 
This can be done in either real space (\eg\ Bond \etal\ 1998)
or 
harmonic space (Challinor \etal\ 2001),
although the real space pointing matrix is more complicated 
if beam asymmetry information is included.

In summary,
if a survey covers the entire celestial sphere it is preferable to 
use full-sky frequency maps. 
However it is possible that small parts of the sky will be missed due
to either the scan strategy (\cf\ Maino \etal\ 1999) 
or hardware problems during the survey itself. 
If this is the case the best unsmoothed map that could be constructed
would be larger than the actual observed region, 
but the errors around the boundary of this area
would be very high. An inferential approach is possible, 
but significant difficulties are encountered, especially in Fourier
space (Challinor \etal\ 2001).
Fortunately, it is probable that both \MAP\ and \Planck\ will produce
full sky maps at several frequencies, which can then be used to
construct maps of the various astrophysical components.

\subsection{Component separation}
\label{section:separate}

The microwave sky consists of several distinct astrophysical components,
as listed in \sect{intro}. 
Fortunately they have sufficiently distinct spectra that they can
be separated using multi-frequency data. 
Given that \MAP\ and \Planck\ will produce maps in 
five and ten bands, respectively, it should be possible to produce
maps of the various components (particularly the \cmb)
that are relatively free of contamination. 
As with map-making, a number of algorithms have been put forward
for this stage of the data processing, 
although the main focus has been on Fourier space methods 
(\eg\ Tegmark \& Efstathiou 1996; Hobson \etal\ 1998; Bouchet \& 
Gispert 1999; Prunet \etal\ 2001; Stolyarov \etal\ 2001). 
Aside from the expected statistical isotropy of the \cmb\ signal,
one of the reasons for this emphasis has been the simplicity
of beam convolution in harmonic space (\sect{convolve}).
This is critical if smoothed frequency maps are used as the 
effective smoothing scale will vary with frequency if the 
telescope is (close to) diffraction limited.
However if unsmoothed maps are used 
real space 
component separation methods
(\eg\ Baccigalupi \etal\ 2000)
must also come into consideration,
the optimal choice of basis being less clear.

One common aspect of all the separation techniques referenced above
is that much of the (prior) information about both signal and noise 
correlations is disregarded in order to render the problem computationally
feasible. 
In real space the correlations between nearby pixels are ignored,
and in Fourier space it is the mode-mode couplings that are neglected.
Surprisingly, these approximations appear to be unimportant in practice --
even the Galactic components have been recovered with striking 
accuracy. 
The most relevant result to this discussion is the all-sky component
separation to $\lmax \simeq 2500$
presented by Stolyarov \etal\ (2001), as it provides clear
evidence that whatever correlations are present in the full-sky harmonic
basis are unimportant -- there are some errors close to the Galactic 
centre, but they are localised, and there is no sign of this 
affecting the reconstruction globally.

If the Galactic plane is removed prior to component separation 
this one troublesome region is no longer present in the analysis,
but new problems arise.
Firstly, smoothed maps (with frequency-dependent beam-widths) 
cannot be used as input data without inducing errors around the edges
of $\cutsphere$ due to the ill-defined nature of convolution on
the cut sky (\sect{convolve}). 
Even if such errors are deemed acceptable 
(\eg\ Prunet \etal\ 2001) or beam-deconvolved maps are used, the
transformation described in 
\eq{aprime_a} completely changes the correlation structure of the
harmonics.
In particular, the signal-signal correlation matrices are
non-diagonal for all components,
including random fields like the \cmb\ (see \sect{powspec}). 
Whereas the couplings between the spherical harmonic coefficients 
can apparently be disregarded, this has not been demonstrated 
for these induced correlations in the orthonormal basis. 
Prunet \etal\ (2001) performed cut-sky component separation 
including them in full, but were thus limited to $\lmax \simeq 500$, 
the computational task being made considerably larger.

In real space the application of a cut is trivial, provided
that beam-deconvolved maps are used, as it simply requires
that pixels in the removed region be ignored. 
Thus the Baccigalupi \etal\ (2000) method should be 
well suited to a cut-sky analysis.

Given that realistic component separation simulations have only 
recently become available, it is likely that important developments 
in this field
will be made in the near future. For the moment, however, it appears
that separation can usefully be performed on either the full or
cut sky without introducing catastrophic errors.
Provided the models of microwave emissions from the
Galactic plane used in the above simulations are sufficiently realistic,
it may thus be
preferable to generate the full-sky maps of the various astrophysical
components, retaining the option of masking unwanted regions at a 
later stage.

\subsection{Power spectrum estimation}
\label{section:powspec}

If the fluctuations in the early universe were the result of 
inflation (\eg\ Linde 1990) then the \cmb\ is expected to be a Gaussian random 
field, the statistical properties of which can be specified 
completely by its angular power spectrum, $C_l$.
Even if this is not the case,
the power spectrum should encode much of the cosmological
information present. 
It is thus unsurprising that, 
as with map-making and component separation (\sects{map} and
\ref{section:separate}, respectively), many different methods of 
power spectrum estimation have been developed
(\eg\ Tegmark 1997; G\'{o}rski 1994; Bond \etal\ 1998; 
Oh \etal\ 1999; Szapudi \etal\ 2001; Wandelt \etal\ 2001;
Hivon \etal\ 2001). 
Further, sky cuts have been incorporated
into many of these algorithms as it seems certain that the strength
of the Galactic microwave emissions will prevent the \cmb\ from ever
being accurately measured in this region. 
Due to the proliferation of papers on this subject, this discussion 
of power spectrum estimation is limited to a description of a
maximum likelihood formalism using the orthonormal basis functions 
described in \sect{orthog}, with reference to how their behaviour
differs in the low- and high-resolution regimes.

\subsubsection{Maximum likelihood formalism}

The most powerful method of power spectrum estimation is 
maximum likelihood (\eg\ Press \etal\ 1992), 
although this has only been implemented 
to \MAP\ resolution to date (Oh \etal\ 1999). 
By invoking Gaussian statistics 
for both the \cmb\ and the noise, 
it is possible to write down the exact likelihood for 
the observed map (\eg\ G\'{o}rski 1994; Borrill 1999).
On the full sky the effective data-vector is $\hat{\vect{s}} =
\vect{s} + \vect{n}$ (\ie\ the estimator for the true sky, of the 
form discussed in \sect{map} and not the quantity being estimated
here) with the non-cosmological $l = 0$ and $l = 1$ modes
removed. The full likelihood is given by 
\[
\frac{{\rmn d}p}{{\rmn d} \hat{\vect{s}}}
=
\frac{1}{\sqrt{(2 \pi)^{i_{\rmn max} - 4} 
\,\,
{\rmn det}(\matr{S} + \matr{N})}}
\exp
\left[
- \frac{1}{2}
\hat{\vect{s}}^{\rmn T}
(\matr{S} + \matr{N})^{-1}
\hat{\vect{s}}
\right],
\]
\begin{equation}
\label{equation:likelihood_fullsky}
\end{equation}
where 
$\matr{S}$ and
$\matr{N}$ are the signal and noise covariance matrices, respectively.
The assumption of Gaussianity implies that 
$S_{i, i^\prime} = \delta_{i, i^\prime} C_{l(i)}$, 
where $C_l$ is the \cmb\ power spectrum
and 
$l(i)$ is defined in \apdx{sphar}.
The form of $\matr{N}$ is determined by a combination of the 
survey method and the data processing up to this point, but is 
unlikely to have the simple form of \eq{covariance} due to 
the imperfect component separation.
The maximum likelihood calculation consists of finding an 
estimator for the underlying power spectrum,
$\hat{C_l}$, such that \eq{likelihood_fullsky} is maximised,
and there are a number of algorithms for finding this quantity
(\eg\ Bond \etal\ 1998; Oh \etal\ 1999).

The maximum likelihood formalism on the cut sky takes the same 
form as on the full sky, 
but with the data-vector $\hat{\vect{s}}^\prime
= \matr{A}^{\rmn T} \hat{\vect{s}}$ and the covariance matrices
suitably transformed to give
(\cf\ G\'{o}rski 1994)
\begin{eqnarray}
\label{equation:likelihood_cutsky}
\frac{{\rmn d}p}{{\rmn d} \hat{\vect{s}}^\prime}
& = &
\frac{1}{\sqrt{(2 \pi)^{i^\prime_{\rmn max} - 4}
{\rmn det}(\matr{S}^\prime + \matr{N}^\prime)}}
\nonumber \\
& \times &
\exp
\left[
- \frac{1}{2}
\hat{\vect{s}}^{\prime \rmn T}
(\matr{S}^\prime + \matr{N}^\prime)^{-1}
\hat{\vect{s}}^\prime
\right],
\end{eqnarray}
with the four modes containing information on the monopole and
dipole (\sect{method}) again excluded.
The signal covariance matrix is subject to 
a simple similarity transform, 
$\matr{S}^\prime = \matr{A}^{\rmn T} \matr{S} \matr{A}$, but
the same is not 
true for the noise covariance matrix as the
noise field is not band-limited
(a fact critical to the use use of a cut-sky analysis at low resolution,
as discussed below).
The coupling of the cut-sky modes makes
the maximisation of \eq{likelihood_cutsky}
non-trivial (\cf\ Oh \etal\ 1999),
even if $\matr{S} + \matr{N}$ is diagonal on the full sky.
Nonetheless it is useful to work under this idealised assumption 
in order to see how the application of the cut ensures that the 
maximum likelihood solution is independent of the Galactic
signal; the manner in which this is achieved is quite different 
in the low- and high-resolution cases.

\subsubsection{Low-resolution analysis}
\label{section:pslowres}

\begin{figure*}
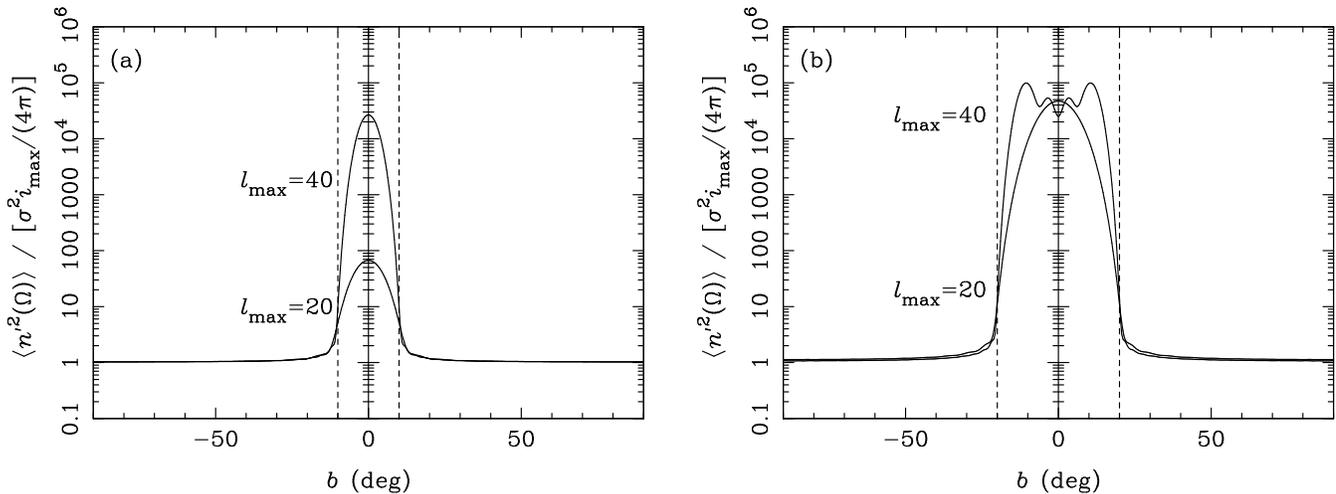

\includegraphics{noise10.ps}
\includegraphics{noise20.ps}
\vspace{\singlefigureheight}
\caption{The variance of a (initially) uniform Gaussian noise field
as a function of latitude,
$b$, after application of the cut-sky orthogonalisation described
in \sect{lowres_harm}.
Constant latitude cuts of $b_{\rmn cut} = 10$ deg (a)
and $b_{\rmn cut} = 20$ deg (b) were applied
(as indicated by the dashed vertical lines)
and results are shown for
$\lmax = 20$ and $\lmax = 40$, as labelled. 
(The oscillations near the peak of the latter curve are indicative 
of the limited accuracy of the decomposition of the ill-conditioned
couling matrix.)}
\label{figure:noise_theta}
\end{figure*}

The effect of a sky cut on power spectrum estimation is 
not entirely obvious in the low-$\lmax$ case in which
the coupling matrix (\sect{coupling}) is invertible. 
The effective band-limit,
produced by the combined effects of the beam and 
noise (\sects{anal} and \ref{section:map}),
means the signal over the whole sky 
(including \eg\ the Galactic plane) is encoded in the cut-sky coefficients.
Thus the application of a cut would be redundant were it not
for the presence of non-band-limited noise which cannot be 
characterised properly a finite harmonic analysis.

Applying the incomplete spherical transform defined in \eq{aprime_def}
to a purely white noise field $n(\skyhat)$
[\ie\ $\langle n(\skyhat) \rangle = 0$
and
$\langle n(\skyhat_1) n(\skyhat_2) \rangle
= \delta(\skyhat_1 - \skyhat_2) \sigma^2$; \cf\ \eq{covariance_real}]
gives cut-sky harmonic coefficients
with $\langle \vect{n}^\prime \rangle = \vect{0}$
and 
\begin{equation}
\matr{N}^\prime = \langle \vect{n}^\prime {\vect{n}^\prime}^{\rmn T} \rangle
= \sigma^2 \tilde{\matr{I}}.
\end{equation}
Projecting back into real space
gives a field $n^\prime(\skyhat)$ 
which satisfies
$\langle n^\prime(\skyhat) \rangle = 0$
and
\begin{equation}
\label{equation:annoying}
\langle n^\prime(\skyhat_1) n^\prime(\skyhat_2) \rangle
= \sigma^2 \vect{Y}^{\rmn T} (\skyhat_1)
\matr{C}^{-1} \vect{Y} (\skyhat_2).
\end{equation}

On the full sky the same procedure (\ie\ a finite spherical 
harmonic analysis followed by a transformation back into real space)
would yield a noise field with covariance structure given by
\begin{eqnarray}
\langle \hat{n} (\skyhat_1) \hat{n} (\skyhat_2) \rangle 
& = & \sigma^2 \vect{Y}^{\rmn T} (\skyhat_1) \vect{Y} (\skyhat_2)
\nonumber \\
& = & \frac{\sigma^2}{4 \pi} \sum_{l = 0}^{\lmax} 
(2 l + 1) P_l(\skyhat_1 \cdot \skyhat_2), 
\end{eqnarray}
where $P_l(x)$ is a Legendre polynomial (\apdx{sphar}).
Taking the limit $\skyhat_2 \rightarrow \skyhat_1$ this implies that
$\langle \hat{n}^2 (\skyhat) \rangle =
i_{\rmn max} \sigma^2 / (4 \pi)$, which represents 
smoothing relative to the original noise field caused by the use of
a finite analysis. 

Whilst this smoothing occurs in
both the cut- and full-sky formalisms,
the presence of $\matr{C}^{-1}$ in the former
case [\eq{annoying}] implies a spatial dependence. 
As can be seen from \fig{noise_theta}, 
the noise in the cut region is greatly increased, 
which is a natural way of formally encoding the qualitative 
fact that, for whatever reason, the data in the cut is contaminated 
by more than just the white noise field.
Thus,
despite the invertibility of the coupling matrix (and the band-limited
cut-sky analysis), the application of a cut has the desired effect
of greatly reducing the impact of any spurious signal, such as the
Galaxy. 
However \fig{noise_theta} also implies that
a similar effect could be achieved without performing a cut
(and hence leaving the signal unchanged),
instead adding a high level of artificial noise in the offending
region(s).
Finally, it is important to note that 
the dual assumptions used in the derivation of \eq{annoying}
-- uniform noise and no beam -- are unrealistic, but the manner in 
which a low-resolution cut-sky analysis works is the same in less
idealised scenarios.

\subsubsection{High-resolution analysis}
\label{section:pshighres}

The high-resolution case is more straightforward,
as the application of the cut results in a data-vector,
$\hat{\vect{s}}^\prime$, which contains little information 
about the removed region. This is quite distinct from the 
low-resolution case discussed above, in that here it is the 
predominantly the signal that is changed, rather than the noise.
That said, the noise close to the boundary of the cut is 
increased in the same manner as explained above. This 
has the same effect as the apodizing function formalism 
described by Tegmark (1997), downweighting points around which
there is not full correlation information. 

Another difference between the low- and high-resolution 
analyses is that 
$\hat{\vect{s}}^\prime$ is smaller than
$\hat{\vect{s}}$, from \sect{eigen}. 
Although this does not result in any significant computational
saving, it serves to emphasise the information loss associated
with removing part of the sky, and is an independent derivation 
of the fact that the uncertainties in the estimated 
power spectrum increase as $4 \pi / \Omega_\cutsphere$
(\cf\ Hobson \& Magueijo 1996; Tegmark 1997).

\section{Conclusions}
\label{section:concs}

The upcoming microwave surveys will require 
a cut-sky analysis to prevent the 
strong Galactic emissions from contaminating 
the \cmb\ signal. 
The spherical harmonics are non-orthogonal on the cut-sphere,
but an orthonormal basis set can be constructed from them 
using \svd-based techniques (\sect{orthog}). 
The application of the resultant conversion matrix to the 
conventional multipoles results in cut-sphere harmonics that
contain only the desired information.
In the low-resolution case the influence of the Galaxy is
reduced by increasing the effective noise in the cut;
in the high-resolution limit the 
orthonormal basis functions can model the infinitely sharp
cut sufficiently well that they have no support in the removed
region.
It is also important to note that the cut should probably only be applied 
after beam-deconvolution has been attempted, as convolution is 
ill-defined on the incomplete sphere. 

The algorithms described here were implemented to 
Legendre multipoles of $\lmax \simeq 2500$
for a constant latitude cut, in
which case the coupling
matrix of the spherical harmonics is block-diagonal.
At present computational limitations make a general
orthogonalisation impractical for $\lmax \ga 200$, 
although there are some possibilities to extend this. 
For instance only $\sim 1$ per cent of the coupling matrix
contains significant information if the cut is well-chosen
(\eg\ rectangular in $\theta$ and $\phi$) and so
sparse matrix techniques should thus allow orthogonalisation to
$\lmax \simeq 1000$ in this case.

Another requirement is orthogonalisation of tensor basis functions
on the incomplete sphere, as both the \MAP\ and \Planck\ satellites
will measure polarization.
The resultant formalism is more complicated, but the same general
principles hold; this issue is explored further by 
Lewis, Challinor \& Turok (2001).

\section*{Acknowledgments}

This paper benefited from useful discussions with several
members of the \Planck\ collaboration,
in particular
Mark Ashdown, 
Fran\c{c}ois Bouchet,
Martin Bucher,
Rob Crittenden,
Jacques Delabrouille,
George Efstathiou,
Krzysztof G\'{o}rski,
Floor van Leeuwen
and Ben Wandelt.
DJM was funded by PPARC.
ADC acknowledges a PPARC Postdoctoral Research Fellowship.
MPH acknowledges a PPARC Advanced Fellowship.

\appendix
\section{Spherical harmonics}
\label{section:sphar}

The spherical harmonics form a complete set of orthonormal basis functions 
over the entire sphere. They are most commonly defined 
as complex functions 
(\eg\ Landau \& Lifshitz 1976; Brink \& Satchler 1993),
but it is more convenient to use real harmonics in this application.
Adapting the notation of G\'{o}rski (1994), the real spherical harmonics are
given by
\begin{equation}
\label{equation:ylmdef}
Y_{l, m}(\skyhat) 
=
Y_{l, m}(\theta, \phi)
=
\lambda_{l,|m|}[\cos(\theta)] s_m(\phi), 
\end{equation}
where $l \geq 0$ and $|m| \leq l$ and 
\begin{equation}
\label{equation:s_m}
s_m(\phi) = \left\{
\begin{array}{lll}
\sqrt{2} \sin(|m| \phi), & {\rmn if} & m < 0, \\
\\
1, & {\rmn if} & m = 0, \\
\\
\sqrt{2} \cos(|m| \phi), & {\rmn if} & m > 0,
\end{array}
\right.
\end{equation}
implying that $\int_0^{2 \pi} s_m(\phi) s_{m^\prime} (\phi) \,
{\rmn d} \phi = 2 \pi \delta_{m, m^\prime}$.
For $0 \leq m \leq l$ and $- 1 \leq x \leq 1$ 
the normalised associated Legendre functions are defined by
\begin{equation}
\label{equation:lambdalmdef}
\lambda_{l,m}(x)
= \sqrt{\frac{2 l + 1}{4 \pi}
\frac{(l - m)!}{(l + m)!}} P_{l,m}(x).
\end{equation}
Hence $\int_{-1}^1 \lambda_{l, m}(x) \lambda_{l^\prime, m} (x)
\, {\rmn d} x = \delta_{l, l^\prime} / (2 \pi)$.
Under the same conditions
the (unnormalised) associated Legendre functions are 
given by\footnote{This definition, with the $(-1)^m$ term, 
correpsonds to that given by 
Abramowitz \& Stegun (1971) and
Gradshteyn \& Ryzhik (2000)
but differs from that used 
by Arfken (1985) and Brink \& Satchler (1993).}
\begin{equation}
\label{equation:plmdef}
P_{l,m}(x) = (- 1)^m \left(1 - x^2 \right)^{m/2}
\frac{{\rmn d}^m}{{\rmn d} x^m}
P_l(x),
\end{equation}
with the Legendre polynomials given by
\begin{equation}
\label{equation:pldef}
P_l(x) = \frac{(-1)^l}{2^l l!}
\frac{{\rmn d}^l}{{\rmn d} x^l}
\left(1 - x^2\right)^l.
\end{equation}

A real field on the sphere, $a(\skyhat)$, can be expanded in terms 
of spherical harmonic coefficients, given by
\begin{equation}
\label{equation:a_lm}
a_{l,m} = \int_{\sphere}
Y_{l, m}(\skyhat) a(\skyhat)
\, {\rmn d}\Omega.
\end{equation}
This can be inverted to give 
\begin{equation}
\label{equation:f_alm}
a(\skyhat) = \sum_{l = 0}^{l_{\rmn max}}
\sum_{m = - l}^{l}
a_{l, m} Y_{l,m} (\skyhat),
\end{equation}
provided that $\lmax \rightarrow \infty$, due 
to the orthonormality of the spherical harmonics on the full sphere:
\begin{equation}
\label{equation:ortho}
\int_{\sphere}
Y_{l,m}(\skyhat) Y_{l^\prime, m^\prime}(\skyhat)
\, {\rmn d}\Omega
= \delta_{l,l^\prime} \delta_{m,m^\prime}.
\end{equation}
If a finite $\lmax$ is used this inversion is no longer possible 
for general $a(\skyhat)$, although it does still hold for band-limited 
functions\footnote{A band-limited function can, by (somewhat circular) 
definition, be constructed from a finite sum over spherical harmonics.}.

Whilst the two indices $l$ and $m$ have quite distinct interpretations
it is convenient to combine them into a single index, $i$, which
allows the definition of vectors $\vect{Y}(\skyhat) = Y_{i(l,m)}(\skyhat)$
and $\vect{a} = a_{i(l,m)}$. Two obvious indexing schemes present 
themselves: grouping in $l$ and $m$. 
The first, as introduced by G\'{o}rski (1994), is natural for power 
spectrum estimation and very simple:
\begin{equation}
i(l, m) = l^2 + l + m + 1.
\end{equation} 
The two `inverses' of this relationship are 
\begin{equation}
l = \integer\, \left[(i - 1)^{1/2} \right]
\end{equation}
and 
\begin{equation}
m = i - (l^2 + l + 1).
\end{equation}
The second choice of ordering
is useful in cases of azimuthal symmetry in which the 
orthogonality expressed in \eq{s_m} is maintained, and 
grouping in $m$ is achieved by defining
\[
i(l, m) =
\left\{
\begin{array}{lll}
l + m + 1 \\
\mbox{} + (l_{\rmn max} + m) (l_{\rmn max} + m + 1)/2,
& {\rmn if} & m \leq 0, \\
\\
l - l_{\rmn max} + (l_{\rmn max} + 1)^2
\\
\mbox{}
- (l_{\rmn max} - m) (l_{\rmn max} - m + 1) / 2,
& {\rmn if} & m > 0.
\end{array}
\right.
\]
\begin{equation}
\label{equation:indexing}
\end{equation}
The `inverses' in this case are given by
\[
m =
\left\{
\begin{array}{lll}
\multicolumn{3}{l}{
\!\!\!\!
\integer \left[- \lmax + \frac{1}{2} \left(\sqrt{8 i + 1} - 3\right)\right],}
\\
\multicolumn{3}{r}{\mathword{if} i \leq (\lmax + 1) (\lmax + 2) / 2,} \\
\\
\multicolumn{3}{l}{
\!\!\!\!
\integer\left[\!\!\left[\lmax - \frac{1}{2} \left\{\sqrt{
8 [(\lmax + 1)^2 - i + 1] + 1} - 3\right\}\right]\!\!\right],} \\
\multicolumn{3}{r}{
\mathword{if} i > (\lmax + 1) (\lmax + 2) / 2}
\end{array}
\right.
\]
\begin{equation}
\end{equation}
and
\begin{equation}
l =
\left\{
\begin{array}{lll}
i - \left[m + 1 \right. \\
\mbox{} \left.
+ (l_{\rmn max} + m) (l_{\rmn max} + m + 1)/2 \right],
& {\rmn if} & m \leq 0, \\
\\
i - \left[- l_{\rmn max} + (l_{\rmn max} + 1)^2 \right.
\\
\mbox{}
\left.
- (l_{\rmn max} - m) (l_{\rmn max} - m + 1) / 2\right],
& {\rmn if} & m > 0.
\end{array}
\right.
\!\!\!\!\!\!\!
\end{equation}
Other indexing schemes have been used in the more specific case 
of simulated {\em Planck} data-sets in which the sky coverage 
is periodic in azimuth (van Leeuwen, private communication),
but are beyond the scope of this paper.

\section{Integration of products of associated Legendre functions}
\label{section:integ}

In \sect{constlat} integrals of the form
\begin{equation}
I_{l, l^\prime, m}(x_1, x_2) = 
\int_{x_1}^{x_2} \lambda_{l,m}(x) \lambda_{l^\prime,m}(x)
\, {\rmn d}x
\end{equation}
arose; here the $\lambda_{l,m}(x)$ are the normalised
associated Ledengre functions, 
defined in \eq{lambdalmdef}, and $m$ is assumed to be non-negative.
These integrals can be evaluated quickly and accurately using a
combination of closed formul\ae\ and recursion relations. 

The associated Legendre functions, $P_{l,m}(x)$ [defined in \eq{plmdef}],
are solutions of the ordinary differential equation (\eg\ Arfken 1985)
\begin{equation}
\label{equation:adjoint}
\frac{{\rmn d}}{{\rmn d}x}
\!\!
\left[
\left(1 - x^2 \right) \frac{{\rmn d} P_{l, m}}{{\rmn d}x}
\right]
+
\left[
l (l + 1) - \frac{m^2}{1 - x^2}
\right] P_{l,m}(x) = 0.
\end{equation}
Multiplying this equation by $P_{l^\prime, m}(x)$ and integrating 
(from $x_1$ to $x_2$) by parts twice yields
\begin{equation}
\label{equation:adjoint_int}
(l - l^\prime) (l + l^\prime + 1) 
\int_{x_1}^{x_2} P_{l, m}(x) P_{l^\prime, x} (x) \, {\rmn d}x
\end{equation}
\[
 = \!\!\! \left. \left[ \!\! \left(1 - x^2 \right) P_{l, m}(x) 
\frac{{\rmn d}P_{l^\prime, m}}{{\rmn d}x}
- \left(1 - x^2 \right) P_{l^\prime, m}(x) 
\frac{{\rmn d}P_{l, m}}{{\rmn d}x}
\right] \! \right|_{x = x_1}^{x = x_2}
.
\]
This is a reflection of the standard result that integrals of 
solutions of a self-adjoint differential equation [as \eq{adjoint} is]
can be expressed as boundary terms (\eg\ Arfken 1985).
The derivatives in \eq{adjoint_int} can be removed by using 
the standard recursion relationship
(\eg\ Gradshteyn \& Ryzhik 2000) 
\begin{equation}
\label{equation:recurs_1}
\left(1 - x^2 \right) \frac{{\rmn d}P_{l,m}}{{\rmn d}x}
= (l + m) P_{l - 1, m}(x) - l x P_{l, m}(x)
\end{equation}
to yield, for $l \neq l^\prime$,
\[
\int_{x_1}^{x_2} P_{l, m}(x) P_{l^\prime, x} (x) \, {\rmn d}x
= \frac{1}{(l - l^\prime) (l + l^\prime + 1)}
\]
\begin{eqnarray}
& \times & 
\left[
(l^\prime + m) P_{l, m}(x) P_{l^\prime - 1, m}(x)
\right. \nonumber \\
& &  
+ (l - l^\prime) x P_{l,m}(x) P_{l^\prime, m}(x)
\nonumber \\
& & \left. \left.
- (l + m) P_{l - 1, m}(x) P_{l^\prime, m}(x)
\right] \right|_{x = x_1}^{x = x_2}.
\end{eqnarray}
Note that the first term must be omitted if $l^\prime = m$ 
and that the third term must be omitted if $l = m$; 
these Legendre functions are implicitly zero from \eq{plmdef}.
Finally, this can be normalised 
according to \eq{lambdalmdef}, giving 
\begin{equation}
\label{equation:closedform}
I_{l, l^\prime, m}(x_1, x_2) = \frac{1}{(l - l^\prime) (l + l^\prime + 1)}
\end{equation}
\begin{eqnarray}
& \times & 
\left[
\sqrt{\frac{2 l^\prime + 1}{2 l^\prime - 1} (l^2 - m^2)}
\lambda_{l, m}(x) \lambda_{l^\prime - 1, m}(x)
\right. \nonumber \\
& & \mbox{} + (l - l^\prime) x \lambda_{l, m}(x) \lambda_{l^\prime, m}(x)
\nonumber \\
& & \left. \left.
 - \sqrt{\frac{2 l + 1}{2 l - 1} (l^2 - m^2)}
\lambda_{l - 1, m}(x) \lambda_{l^\prime, m}(x)
\right] \right|_{x = x_1}^{x = x_2}.
\nonumber 
\end{eqnarray}
An alternative derivation of this result was presented by 
Wandelt \etal\ (2001); it is also in principle equivalent to
Eq.\ 5.9(13) of Varshalovich, Moskalev \& Khersonskii (1988),
but their application of \eq{recurs_1} is in error.

For the case $l = l^\prime$ a recursion relation is required,
starting with $l = m$.
Combining \eqs{lambdalmdef} and (\ref{equation:plmdef}),
\begin{equation}
\label{equation:lambda_mm}
\lambda_{m,m}(x) = 
(- 1)^m \sqrt{\frac{2 m + 1}{4 \pi}}
(2 m - 1)!! \left(1 - x^2\right)^{m/2},
\end{equation}
where $n!! = 1 \times 3 \times \cdots \times (n -2) \times n$ for odd $n$.
Integrating by parts and using \eq{lambda_mm} again gives
\begin{equation}
\label{equation:i_mm}
I_{m,m,m}(x_1, x_2) 
\end{equation}
\[
\mbox{}
= 
\left\{
\begin{array}{lll}
\frac{x_2 - x_1}{4 \pi}, & {\rmn if} &  m = 0, \\
\\
I_{m - 1, m - 1, m - 1}(x_1, x_2)
+
\frac{
\left. x \lambda_{m, m}^2(x) \right|_{x_1}^{x_2}
}{2 m + 1},
 & {\rmn if} & m > 0.
\end{array}
\right.
\]

Moving to $l = m + 1$, the standard relationship 
(\eg\ Gradshteyn \& Ryzhik 2000) that
\begin{equation}
\lambda_{m + 1, m}(x) = (2 m + 3) x \lambda_{m, m}(x)
\end{equation}
combines with \eq{lambda_mm} to give
\begin{eqnarray}
\label{equation:i_mpmp}
I_{m + 1, m + 1, m} (x_1, x_2) 
& = & I_{m, m, m} (x_1, x_2) \nonumber \\
& - &
\frac{2 m + 2}{2 m + 3}
\left. x \, \lambda^2_{m + 1, m + 1}(x) \right|_{x_1}^{x_2},
\end{eqnarray}
where the first term is given in \eq{i_mm}.

The last step is to derive a recursion relation relating 
$I_{l, l, m}(x_1, x_2)$ 
to
$I_{l - 1, l - 1, m}(x_1, x_2)$
and
$I_{l - 2, l - 2, m}(x_1, x_2)$.
Eq.\ (C22) of Wandelt \etal\ (2001) gives a four-term recursion 
to obtain $I_{l, l^\prime, m}(x_1, x_2)$; it can be applied 
successively (once swapping $l$ and $l^\prime$) to obtain
\[
I_{l, l, m}(x_1, x_2) =
-
\frac{2 m^2 - 2 l^2 + 2 l - 1}{(l - m)(l + m)} 
I_{l - 1, l - 1, m}(x_1, x_2)
\]
\begin{eqnarray}
\label{equation:i_ll}
& - & \frac{(l - 1 - m) (l - 1 + m)}{(l - m)(l + m)} 
I_{l - 2, l - 2, m}(x_1, x_2)
\nonumber \\
& - & \sqrt{\frac{2 l - 1}{2 l + 1} \frac{1}{(l - m)(l + m)}}
\nonumber \\
& & \times 
\left.
\left(1 - x^2 \right) \lambda_{l, m}(x) \lambda_{l - 1, m}(x)
\right|_{x = x_1}^{x = x_2}
\nonumber \\
& + & 
\sqrt{\frac{2 l - 1}{2 l - 3} 
\frac{(l - 1 - m)(l - 1 + m)}{(l - m)^2(l + m)^2}} 
\nonumber \\
& & \times 
\left.
\left(1 - x^2 \right) \lambda_{l - 1, m}(x) \lambda_{l - 2, m}(x)
\right|_{x = x_1}^{x = x_2}.
\nonumber 
\end{eqnarray}

In summary, \eq{closedform} can be used to evaluate all
$I_{l, l^\prime, m}(x_1, x_2)$ for which $l \neq l^\prime$,
and \eqs{i_mm}, (\ref{equation:i_mpmp}) and 
(\ref{equation:i_ll}) combine to give
all $I_{l, l, m}(x_1, x_2)$ recursively.

\section{Treatment of non-cosmological modes}
\label{section:noncosmol}

All the cosmological information encoded in the \cmb\ is expected to be
contained in the $l \geq 2$ modes;
the $l = 0$ mode in an isotropic universe can be normalised 
arbitrarily
and the $l = 1$ modes can be set to zero by adopting an appropriate 
reference frame.
Nonetheless, observations of the 
microwave sky will yield non-zero monopole and dipole values 
for a number of reasons (\eg\ the observer's motion; Galactic emission;
extra-Galactic point-sources).
Hence these low-order modes must be included in the analysis of \cmb\
data, but should be kept separate from the cosmological modes, as 
is naturally the case if spherical harmonic coefficients are used to 
describe the data.
It is also important to note that the properties of the basis functions
themselves are unimportant -- the essential requirement is that 
only four of the cut-sky harmonic coefficients contain information
on the unwanted modes. 

The method of orthogonalisation summarised in
\eqs{b_general}, 
(\ref{equation:coupling_decomp}) 
and 
(\ref{equation:b_decomp})
does not explicitly impose any particular structure on the conversion
matrix, $\matr{A}$ [which relates harmonic coefficients on the 
incomplete sphere to those on the full sphere by
$\vect{a}^\prime = \matr{A}^{\rmn T} \vect{a}$; \eq{aprime_a}]. 
The non-cosmological modes are kept separate from the cosmological
modes if the first four columns of $\matr{A}^{\rmn T}$ 
have only zeros from the fifth row on, assuming the full-sky harmonic
coefficients are indexed using $l$-ordering (\apdx{sphar}).
This is achieved naturally if $\matr{A}^{\rmn T}$ is constructed to be upper
triangular, as in the case of the Cholesky decomposition described
in \sect{lowres_orthog}. The other 
decomposition methods discussed in \sect{method}
do not share this property, and so the resultant conversion matrices
must be adjusted explicitly. 

One way of achieving this is to use a partial Householder
transform (\eg\ Press \etal\ 1992).
The last $i^\prime_{\rmn max} - i$ elements of the
$i$th column of a general 
$i^\prime_{\rmn max} \times i_{\rmn max}$ matrix 
can be set to zero by the transformation 
$\matr{M}^\prime = \matr{P}_i \matr{M}$, 
with the orthogonal Householder matrix defined by 
\begin{equation}
\matr{P}_i = 
\tilde{\matr{I}} - 2 \frac{
\vect{m}_i \vect{m}_i^{\rmn T}}{\vect{m}_i^{\rmn T} \vect{m}_i},
\end{equation}
where $\vect{m}_i$ is given by
\begin{equation}
(m_i)_j = 
\left\{
\begin{array}{lll}
0, & {\rmn if} & j < i, \\
& & \\
M_{j,i} + \sqrt{\sum_{k = i}^{i^\prime_{\rmn max}} M_{k,i}^2},
& {\rmn if} & j = i, \\
& & \\
M_{j,i} ,
& {\rmn if} & j > i,
\end{array}
\right. 
\end{equation}
and $i \leq \min (i^\prime_{\rmn max}, i_{\rmn max}$) is assumed.
Provided that the Householder matrix applied to $\matr{B}$
is that generated from $\matr{A}^{\rmn T}$, 
the transformations
${\matr{A}^\prime}^{\rmn T} = \matr{P}_i \matr{A}^{\rmn T}$
and
$\matr{B}^\prime = \matr{P}_i \matr{B}$
leave \eqs{coupling_decomp} and (\ref{equation:b_decomp}) unaffected
as $\matr{P}_i$ is orthogonal by construction.
Applying $\matr{P}_1$,
$\matr{P}_2$,
$\matr{P}_3$
and then
$\matr{P}_4$
to the successively updated $\matr{A}^{\rmn T}$ 
ensures that 
the $l = 0$ and $l = 1$ modes influence only the first four
cut-sky harmonic coefficients, as required.
This procedure could be continued, moving $\matr{A}^{\rmn T}$ 
successively closer to upper triangular form, although this
cannot be achieved in full as 
$\matr{A}^{\rmn T}$ has more columns than rows.

Special mention must be made of the constant latitude cut case,
the symmetry of which can only be utilised if the spherical harmonics
are indexed using $m$-ordering (\apdx{sphar}). In this case only
the $m = 0$ and $m = \pm 1$ blocks have any contribution from the 
monopole or dipole, and each can be treated separately. 
Further, the ordering within these blocks is such that the 
non-cosmological modes are in the first rows, and so the above 
algorithm can be applied to each of three blocks as is. 
The only slight inconvenience is that it is no longer the first
four cut-sky modes that contain the non-cosmological information,
and the relevant modes must be flagged explicitly.

\bsp
\label{lastpage}
\end{document}